\documentclass[%
 aip,
 amsmath,amssymb,
 reprint,%
]{revtex4-1}

\usepackage{graphicx}% Include figure files
\usepackage{dcolumn}% Align table columns on decimal point
\usepackage{bm}% bold math
\usepackage[utf8]{inputenc}
\usepackage[T1]{fontenc}
\usepackage{mathptmx}

\DeclareMathAlphabet{\mathcal}{OMS}{cmsy}{m}{n}

\begin{document}

\preprint{AIP/123-QED}

%\title{Kinetic Entropy as a Diagnostic in Particle-in-Cell Simulations
%  of Heliospheric, Planetary, and Astrophysical Plasmas}
\title{Decomposition of Plasma Kinetic Entropy into Position and Velocity Space and the Use of Kinetic Entropy in Particle-in-Cell Simulations}
% Force line breaks with \\

\author{Haoming Liang}
\email{haoming.liang@mail.wvu.edu.}
\author{Paul A.~Cassak} 
\affiliation{Department of Physics and Astronomy, West Virginia University, Morgantown, WV 26506, USA} 
\author{Sergio Servidio} 
\affiliation{Dipartimento di Fisica, Universit\`a della Calabria, I-87036 Cosenza, Italy}
\author{Michael A.~Shay}
\affiliation{Department of Physics and Astronomy, University of Delaware, Newark, DE 19716, USA}
\author{James F.~Drake, Marc Swisdak}
\affiliation{Department of Physics and Institute for Research in Electronics and Applied Physics, University of Maryland, College Park, MD 20740, USA}
\author{Matt R.~Argall} 
\affiliation{Physics Department and Space Science Center, Morse Hall, University of New Hampshire, Durham, NH 03824, USA}
\author{John C.~Dorelli}
\affiliation{NASA Goddard Space Flight Center, Greenbelt, Maryland 20771, USA}
\author{Earl E.~Scime}
\affiliation{Department of Physics and Astronomy, West Virginia University,
 Morgantown, WV 26506, USA}
\author{William H.~Matthaeus}
\affiliation{Department of Physics and Astronomy, University of Delaware, Newark, DE 19716, USA}
\author{Vadim Roytershteyn}
\affiliation{Space Science Institute, Boulder, Colorado 80301, USA}
\author{Gian Luca Delzanno}
\affiliation{T-5 Applied Mathematics and Plasma Physics Group, Los Alamos National Laboratory, Los Alamos, NM 87545, USA}

\date{\today}% It is always \today, today,
             %  but any date may be explicitly specified

\begin{abstract}
  We describe a systematic development of kinetic entropy as a diagnostic in fully kinetic particle-in-cell (PIC) simulations and use it to interpret plasma physics processes in heliospheric, planetary, and astrophysical systems. First, we calculate kinetic entropy in two forms -- the ``combinatorial'' form related to the logarithm of the number of microstates per macrostate and the ``continuous'' form related to $f \ln f$, where $f$ is the particle distribution function. We discuss the advantages and disadvantages of each and discuss subtleties about implementing them in PIC codes. Using collisionless PIC simulations that are two-dimensional in position space and three-dimensional in velocity space, we verify the implementation of the kinetic entropy diagnostics and discuss how to optimize numerical parameters to ensure accurate results. We show the total kinetic entropy is conserved to three percent in an optimized simulation of anti-parallel magnetic reconnection. Kinetic entropy can be decomposed into a sum of a position space entropy and a velocity space entropy, and we use this to investigate the nature of kinetic entropy transport during collisionless reconnection. We find the velocity space entropy of both electrons and ions increases in time due to plasma heating during magnetic reconnection, while the position space entropy decreases due to plasma compression. This project uses collisionless simulations, so it cannot address physical dissipation mechanisms; nonetheless, the infrastructure developed here should be useful for studies of collisional or weakly collisional heliospheric, planetary, and astrophysical systems. Beyond reconnection, the diagnostic is expected to be applicable to plasma turbulence and collisionless shocks.
\end{abstract}

\maketitle

\section{Introduction}

Dissipation of energy in nearly collisionless plasmas is a key
component of understanding many fundamental plasma processes, such as
magnetic reconnection, plasma turbulence, and collisionless shocks.
In magnetic reconnection, dissipation can change magnetic topology
 \citep{Hesse11,Cassak16} and may play a role in
thermalizing plasma in the exhausts \citep{Drake06}.  In
plasma turbulence, dissipation at kinetic scales is required to
terminate the energy cascade 
  \citep{Cranmer02,Parashar09}.  A number of mechanisms for this
conversion in weakly collisional plasmas have been discussed,
including resonant and non-resonant wave-particle interactions and
dissipation in coherent structures ({\it i.e.,} intermittency) such as
through reconnection \citep{Howes18}.  In collisionless shocks,
dissipation is necessary to convert the upstream plasma bulk flow
energy into thermal energy  \citep{Krall97}.  These three
fundamental processes underlie a staggering array of important
applications in heliospheric, planetary, and astrophysical sciences,
including supernova shocks \citep{Reynolds08}, astrophysical jets
\citep{Beall14}, pulsar winds \citep{Gaensler06}, interstellar shocks
\citep{Draine93}, shocks in galaxy cluster mergers \citep{Ensslin97},
solar eruptions \citep{Priest02}, coronal heating \citep{Klimchuk06},
solar wind turbulence \citep{Gosling07}, solar wind-magnetosphere
coupling and magnetospheric storms and substorms 
  \citep{Kivelson95}, and planetary shocks \citep{Tsurutani13}.

The study of dissipation is at the forefront of research in these
processes and settings, but it has been challenging to study it
observationally, experimentally, numerically, and theoretically \citep{Cassak16,Howes18}.  Recently, dissipation has
become more accessible to study numerically through increases in
computer power and observationally through the development of high
cadence satellite measurements.  For example, the primary objective of the Magnetospheric
Multiscale (MMS) mission \citep{Burch16} is dissipation accompanying
reconnection \citep{Burch16b}, and it has
also been used to study magnetosheath turbulence \citep{Servidio17,Chen19}
and the bow shock \citep{Chen18}.  Studying dissipation in solar wind
turbulence would have been a key goal of the Turbulence Heating
ObserveR (THOR) mission \citep{Vaivads16}.

From a theoretical perspective, there have been efforts to identify
regions where dissipation occurs.  These measures have had some
success in identifying the electron diffusion region (EDR)
\citep{Shay07} of magnetic reconnection 
  \citep{Zenitani11,Swisdak16,Burch16b,Ashour-Abdalla16} and
dissipation in reconnection exhausts \citep{Sitnov18}, and dissipation
in plasma turbulence \citep{Wan16,Yang17a,Yang17b}.
However, it is not clear which, if any, uniquely identifies genuine
dissipation.

The present study is based on the premise that entropy is a natural
candidate to identify and quantify dissipation.  Entropy in a closed
system is conserved in the absence of dissipation and monotonically
increases when dissipation is present 
  \citep{Boltzmann77,Bellan08}. Here, we interpret ``dissipation'' as
a process that causes a total entropy increase in a closed, isolated
system.

The fluid (thermodynamic) form of the entropy per particle for an
isotropic plasma is related to $p/\rho^\gamma$, where $p$ is the
(scalar) pressure, $\rho$ is the mass density, and $\gamma$ is the
ratio of specific heats.  This quantity has been studied in various
settings for a long time.  For example, stability of Earth's
magnetotail plasma sheet to the interchange instability is governed by
fluid entropy
\citep{Erickson80,Borovsky98,Kaufmann06,Wolf06,Birn09,Wolf09,Johnson09,Wang09,Sanchez12,Liu14}.
Fluid entropy was specifically investigated in the context of magnetic
reconnection, finding that it is conserved very well in
magnetohydrodynamic (MHD) and particle-in-cell (PIC) simulations of
reconnecting flux tubes \citep{birn05a,Birn06}.  Fluid entropy has
been used to identify non-adiabatic heating during reconnection
\citep{Hesse09,Ma14}.  \citet{Lyubarsky01} used fluid entropy in their
study of reconnection in pulsar winds.  \citet{Rowan17} subtracted
adiabatic heating from measured heating in the exhaust of a
reconnection event in PIC simulations to find the leftover
non-adiabatic contribution.  A similar approach was used to study
entropy production in collisionless shocks in PIC simulations
\citep{Guo17,Guo18}.

Many heliospheric, planetary, and astrophysical settings are only
weakly collisional, so the fluid approximation may or may not be
applicable.  Instead, a kinetic approach is likely necessary in such
settings, especially in regions with fine-scale spatial or temporal
structures.  We follow the convention by \citet{Kadanoff14} and refer
to the version of entropy in kinetic theory as ``kinetic entropy''.
The theory will be reviewed in Appendix~\ref{appendix:background}.

Kinetic entropy has been a useful diagnostic in studies using the
gyrokinetic model.  In this model, the second order perturbed
distribution function is related to the perturbed kinetic entropy
\citep{Krommes94,Schekochihin09} and the kinetic entropy production
rate is related to the heating rate \citep{Howes06}.  Using
gyrokinetic and related models, energy dissipation and plasma heating
have been studied in simulations of magnetic reconnection
\citep{Loureiro13,Numata15} and plasma turbulence
\citep{Watanabe04,Tatsuno09,TenBarge11,Nakata12,TenBarge13,Told15,Li16,Klein17,Grovselj17}.
Kinetic entropy has also been investigated in studies of turbulence
using the Vlasov-hybrid (Vlasov ions, fluid electrons) approach
\citep{Cerri18} and in shocks \citep{Margolin17}.

Meanwhile, the investigation of kinetic entropy in fully kinetic
plasma systems, {\it i.e.,} without any degrees of freedom integrated
out, has been carried out in some observational and theoretical
studies.  Observational data was used to study kinetic entropy in
Earth's plasma sheet \citep{Kaufmann09,Kaufmann11} and Earth's bow
shock \citep{Parks12}.  Dynamics of the magnetosphere was investigated
using various entropy measures from statistics \citep{Balasis09}.
Generalizations of kinetic entropy to kappa-distributions in the solar
wind have been studied \citep{Leubner04}.  The permutation entropy was
used to analyze solar wind turbulence \citep{Olivier18}.  The entropy
production in a kinetic-based fluid closure \citep{Hammett90} was
recently investigated \citep{Sarazin09}.  Kinetic mechanisms for the
increase of entropy have been discussed for reconnection with an
out-of-plane (guide) magnetic field \citep{Hesse17}.  A recent model
of the turbulent cascade employs the kinetic entropy in a
renormalization group approach \citep{Eyink18}.  However, we are not
aware of any studies calculating kinetic entropy from first principles
in fully kinetic PIC simulations.

There are challenges to use entropy as a diagnostic in a real
system. First, the entropy can vary due to inhomogeneous plasma
parameters, such as density and temperature, but mere convection
should not be mistaken for dissipation.  Moreover, equating an entropy
increase with dissipation requires a closed system, but naturally
occurring systems tend not to be closed.  Despite these challenges the
present approach is based on the view that studying entropy in fully
kinetic models (from collisionless to collisional) in closed systems
is useful to understand entropy production.  The insights gained can
be applied to understanding dissipation in real systems.  Therefore,
we argue that kinetic entropy can be a useful measure in collisionless
systems, and can be crucial in collisional systems to identify
dissipation. This is especially the case in the modern age of
observational assets like MMS that measure particle distribution
functions with a cadence of a fraction of a second and with high
resolution in velocity space.

In this work, we describe a systematic development of kinetic entropy
as a diagnostic in fully kinetic PIC simulations and investigate some
of its uses to interpret plasma physics processes in heliospheric,
planetary, and astrophysical systems.  We implement two forms of
kinetic entropy \citep{Boltzmann77,Planck06} in our PIC code, the
``combinatorial'' and ``continuous'' forms.    \citep{Mouhot11,Goldstein04}  We use the kinetic
  entropy diagnostic on a two-dimensional in position space,
  three-dimensional in velocity space collisionless PIC simulation of
  antiparallel magnetic reconnection, though we expect it will be
  equally useful for simulations of plasma turbulence and
  collisionless shocks.  Here, we summarize the new numerical and
  physical contributions resulting from this study:
\begin{enumerate}
\item We perform the first implementation (that we are aware) of
  the direct calculation of the combinatorial kinetic entropy in a PIC
  simulation, and provide a definitive assessment of its advantages
  and disadvantages relative to the more standard continuous kinetic
  entropy form.
\item We perform a careful validation of the kinetic entropy
  diagnostics as a function of numerical parameters, which is
  important to ensure proper application of this approach in future
  studies of reconnection or other applications.  The discussion
  includes how to choose the velocity space grid scale and the number
  of macro-particles per grid cell $PPG$. We point out that
  macro-particles (also known as super-particles) in a PIC simulation
  represent a large number of actual particles in the system being
  simulated, and this needs to be properly accounted for to compare to
  observations or experiments.
\item We show the kinetic entropy increases by only 3\% in a
  carefully constructed collisionless PIC simulation of magnetic
  reconnection. This gives the first estimate that we are aware of the
  fidelity one can expect from a collisionless PIC simulation in
  conserving kinetic entropy. The impact of this result on physics is
  that it shows it will be possible to include collisions into a PIC
  code and expect to be able to resolve its effect on the production
  of entropy through irreversible collisional processes.  This is
  crucial for PIC studies of irreversible dissipation (which is a
  topic of future work).
\item We show that kinetic entropy is not reliably produced in
  simulations with a low number of particles per grid cell. We confirm
  simulations with a reduced number of particles can reproduce
  macroscopic quantities like the reconnection rate, but it may
  (depending on the PIC algorithm) give unphysical results for
  dissipation. This suggests caution is needed for low macro-particle
  per grid cell simulations on matters of kinetic entropy production,
  including particle acceleration and plasma heating. The present
  study provides a blueprint for how studies with a low number of
  particles per grid can determine if their numerics are impacting
  their physical results.
\item We decompose the total kinetic entropy into the sum of a
  position space and velocity space kinetic entropy. That this
  decomposition is possible seems to have been known previously in
  mathematical applications of plasma physics including Landau
  damping \citep{Mouhot11,Goldstein04}, but to our knowledge
  this has not been exploited in applications to magnetized physical
  processes like magnetic reconnection (or turbulence or shocks).
  There are significant reasons this contribution is important to
  studies of entropy and dissipation. We show that this decomposition
  is helpful to understand the dynamics. For both electrons and ions,
  the position space entropy decreases in time during reconnection,
  while the velocity space entropy increases.  This result has a clear
  physical interpretation, as the heating of particles leads to an
  increase in temperature and therefore an increase in velocity space
  entropy, while the compression of upstream particles into the
  current sheet and the magnetic islands leads to a decrease in
  position space entropy.  Therefore, in collisionless systems in
  which total kinetic entropy is conserved, there is a conversion
  between the two types of kinetic entropy.  This result is
  potentially important for observational studies of kinetic entropy.
  It reveals that an increase in the local velocity space kinetic
  entropy need not be associated with dissipation, as it also includes
  contributions from reversible energy conversion due to compression.
  Thus, caution must be employed when studying velocity space kinetic
  entropy.

Another reason decomposing kinetic entropy into position and velocity
space contributions is useful is that in nearly collisionless plasmas
of heliophysical, astrophysical, and planetary interest, particle
distributions can become strongly non-Maxwellian.  The decomposition
of a distribution into a thermal and non-thermal part is not possible
for such complicated distributions.  In a closed system that does not
include collisions, the conservation of total kinetic entropy implies
that any increase in velocity space entropy is balanced by an equal
decrease to the position space entropy, and vice versa. In a closed
collisional system, the two will not be balanced, and the net change
of kinetic entropy gives a measure of the rate of dissipation. An
example of a use of this is that one can tell by comparing the
position and velocity space entropy what portion of the increase in
velocity space kinetic entropy is reversible (the part that goes to
position space kinetic entropy) and what portion is irreversible. This
can be done from a calculation of the distribution function as a
whole, without having to break it up into a thermal and non-thermal
part, so it even works for distributions that are strongly non-
Maxwellian.
\end{enumerate}

It is worth noting that in the present study we develop a framework
and perform a preliminary study, but we do not address the physical
cause of dissipation because its presence in these simulations is
purely numerical.  One can show analytically that kinetic entropy
increases only in the presence of collisions \citep{Bellan08}.  Since
we use a collisionless PIC code for this study, the small kinetic
entropy production we detect is due to numerical effects.  We leave
studies of mechanisms of dissipation for future work using a
collisional PIC model.

This paper is organized as follows: in Sec.~\ref{sec:theory}, 
we briefly list the forms of kinetic entropy that we investigate.  
The existing theory of kinetic entropy including the fact that 
kinetic entropy can be decomposed into position and velocity space entropies 
is reviewed in Appendix~\ref{appendix:theory}.  
Appendix~\ref{appendix:implement} contains a thorough discussion of implementing 
the kinetic entropy diagnostic into PIC codes. 
Section~\ref{sec:simulations} describes the setup of the simulations
we employ.  Section~\ref{sec:results} shows the simulation results,
including a discussion of how to choose the diagnostic and simulation
parameters to achieve robust results and a discussion of using kinetic
entropy to obtain physical insights.  Finally, conclusions,
applications, and future work are discussed in Sec.~\ref{sec:conc1}.

\section{Kinetic Entropies in This Study}
\label{sec:theory}
In this section, we review the forms of kinetic entropy that we
calculate in this study.  The detailed derivation and discussion of
the kinetic entropy expressions are given in
Appendix~\ref{appendix:theory}.

The ``combinatorial Boltzmann entropy'' $\mathcal{S}$ is defined in
Eq.~(\ref{eq:lngamma1}) as
%\begin{linenomath*}
\begin{equation}
\mathcal{S}=k_B\left[\ln N!-\sum_{j,k}\ln N_{jk}!\right], \label{eq:lngamma1_sec2}
\end{equation}
%%\end{linenomath*}
where $N_{jk}$ is the number of particles in phase space bin spanning
$(\vec{r}_j,\vec{v}_k) \rightarrow (\vec{r}_j+\Delta
\vec{r},\vec{v}_k+\Delta \vec{v})$ and $N = \sum_{j,k}N_{jk}$ is the
total number of particles. Since the total number of particles in a
closed system is fixed, percentage changes in entropy are calculated
based solely on the second term.

By using Stirling's approximation and ignoring constant terms, one
obtains the ``continuous Boltzmann entropy'' $S$ in
Eq.~(\ref{eq:stirling_f2}) as
%\begin{linenomath*}
\begin{equation}\label{eq:stirling_f2_sec2}
S = - k_B \int d^3{r}d^3v f(\vec{r},\vec{v}) \left[\ln
  f(\vec{r},\vec{v}) \right],
\end{equation}
%%\end{linenomath*}
where $f(\vec{r},\vec{v})$ is the distribution function at position
$\vec{r}$ and velocity $\vec{v}$ in phase space. The continuous
Boltzmann entropy per unit volume, {\it i.e.,} the continuous
Boltzmann entropy density $s(\vec{r})$, is defined in
Eq.~(\ref{eq:entropy_density}) as
%\begin{linenomath*}
\begin{equation}
s(\vec{r}) = - k_B \int d^3v f(\vec{r},\vec{v}) \left[\ln
  f(\vec{r},\vec{v}) \right]. \label{eq:entropy_density_sec2}
\end{equation}
%%\end{linenomath*}
The continuous Boltzmann entropy density $s_M(\vec{r})$ for a 3D
drifting Maxwellian velocity distribution in local thermodynamic
equilibrium (LTE) for a species of mass $m$, number density
$n(\vec{r})$, bulk flow velocity $\vec{u}(\vec{r})$, and temperature
$T(\vec{r})$, with $f(\vec{r},\vec{v}) = f_{M} = n(\vec{r})[m/2 \pi
  k_B T(\vec{r})]^{3/2} e^{-m [\vec{v} - \vec{u}(\vec{r})]^2 / 2 k_B
  T(\vec{r})}$, follows directly. The result is in
Eq.~(\ref{eq:entropy_density_maxwell}):
%\begin{linenomath*}
\begin{equation}\label{eq:entropy_density_maxwell_sec2}
s_M(\vec{r}) = \frac{3}{2}k_B n(\vec{r}) \left[ 1+\ln \left(\frac{2\pi
    k_B T(\vec{r})}{m n^{2/3}(\vec{r})}\right) \right]. 
\end{equation}
%\end{linenomath*}
We use Eq.~(\ref{eq:entropy_density_maxwell_sec2}) to validate the
implementation of the kinetic entropy diagnostic in our PIC code.

Both the combinatorial and continuous kinetic entropies can be
decomposed into a sum of a position space entropy and a velocity space
entropy. The derivation and discussion of the physical meaning of
these two terms are reviewed in Appendix~\ref{appendix:pos_vel}. We
define the combinatorial position space entropy
$\mathcal{S}_{\text{position}} $ and velocity space entropy
$\mathcal{S}_{\text{velocity}}$ in Eqs.~(\ref{eq:ent_space})
and~(\ref{eq:ent_vel}) as
%\begin{linenomath*}
\begin{equation}\label{eq:ent_space_sec2}
\mathcal{S}_{\text{position}}=k_B\left[\ln N!-\sum_{j}\ln N_j!\right],
\end{equation}
%\end{linenomath*}
%\begin{linenomath*}
\begin{equation}
\mathcal{S}_{\text{velocity}} = \sum_j k_B\left[\ln
  N_j!-\displaystyle\sum_{k}\ln N_{jk}!\right], \label{eq:ent_vel_sec2}
\end{equation}
%\end{linenomath*}
where $N_j$ is the total number of particles in the $j$th spatial bin
by summing $N_{jk}$ only over velocity space. The continuous
  position space entropy $S_{\text{position}}$ and velocity space
  kinetic entropy $S_{\text{velocity}}$ are expressed in
Eqs.~(\ref{eq:space_stirling_lim})-(\ref{eq:svelocity}) as
%\begin{linenomath*}
\begin{eqnarray}
%\\ \text{}_{\text{}}\text{} & \text{} & \nonumber \\
%\textit{S}_{\text{position}} & \text{=}  &
S_{\text{position}} & =  &
k_B\left[N\ln\left(\frac{N}{\Delta^3r}\right) \right. \nonumber \\ &
  -& \left. \int d^3r n(\vec{r})\ln
  n(\vec{r})\right], \label{eq:space_stirling_lim_sec2}
\\
\textit{S}_{\text{velocity}} & \text{=}  &
%\\ S_{\text{velocity}} & = & 
\int d^3r s_{\text{velocity}}(\vec{r}),
\\ s_{\text{velocity}}(\vec{r}) & = &
k_B\left[n(\vec{r})\ln\left(\frac{n(\vec{r})}{\Delta^3v}\right)
  \right.  \nonumber \\ & -& \left. \int d^3v f(\vec{r},\vec{v})\ln
  f(\vec{r},\vec{v})\right], \label{eq:svelocity_sec2}
\end{eqnarray}
%\end{linenomath*}
where $\Delta^3r$ and $\Delta^3v$ are the volumes of the bins in
position and velocity space, respectively , and
  $s_{\text{velocity}}(\vec{r})$ is the continuous velocity space
  kinetic entropy density whose spatial integral gives
  $S_{\text{velocity}}$.  While it is possible in principle to define
  a continuous position space kinetic entropy density, it is not
  unique and it does not have a physical interpretation as the
  permutation of particles in position space, so we do not define a
  position space kinetic entropy density.  Rather, we point out that
  the first term in Eq.~(\ref{eq:space_stirling_lim_sec2}) is a
  constant, so the time evolution of the position space kinetic
  entropy is solely determined by the spatial integral of $-n\ln n$.

Details about how to implement kinetic entropy diagnostics into a PIC
code are discussed in Appendix~\ref{appendix:implement}. The
discussions include the importance of the actual number of particles
per macro-particle, binning particles in phase space, obtaining the
distribution function and kinetic entropies, and a comparison between
combinatorial and continuous Boltzmann entropies.

\section{Simulations}
\label{sec:simulations}

Simulations are carried out using the {\sc p3d} code \citep{Zeiler02},
though we expect the diagnostic and analysis would be possible with
any explicit PIC code.  The code uses the relativistic Boris particle
stepper \citep{Birdsall04} for the particles and trapezoidal leapfrog
\citep{Guzdar93} on the electromagnetic fields, with the fields
allowed to have a smaller time step than the particles (half as big
for our simulations).  The divergence of the electric field is cleaned
(every 10 particle time steps unless otherwise noted for our
simulations) using the multigrid approach \citep{Trottenberg00}.
Boundary conditions in every direction are periodic.  The
normalization is based on an arbitrary magnetic field strength $B_0$
and density $n_0$.  Spatial and temporal scales are normalized to the
ion inertial length $d_i=c/\omega_{pi}$ and the ion cyclotron time
$\Omega_{ci}^{-1}$, respectively, where
$\omega_{pi}=\sqrt{n_0e^2/\epsilon_0m_i}$ is the ion plasma frequency
and $\Omega_{ci}=eB_0/m_i$ is the ion cyclotron frequency based on
$n_0$ and $B_0$.  Thus, velocities are normalized to the Alfv\'{e}n
velocity $v_A=d_i \Omega_{ci}$.  Electric fields are normalized to
$v_A B_0$.  Pressures and temperatures are normalized to $B_0^2 /
\mu_0$ and $m v_A^2/k_B$, respectively.  Entropies are normalized to
Boltzmann's constant $k_B$, though see Appendix~\ref{appendix:calcrealunits}
for a discussion of the units of the continuous Boltzmann entropy.

For simplicity in this initial study, we only consider 2D in position
space, 3D in velocity space simulations of symmetric anti-parallel
magnetic reconnection.  The simulation domain is $L_x \times L_y=51.2
\times 25.6$.  A double current sheet initial condition is used, with
magnetic field given by $B_x(y)= \{\tanh [(y-3L_y/4)/w_0] -
\tanh[(y-L_y/4)/w_0] + 1\}$, where $w_0=0.5$ is the initial
half-thickness of the current sheet.  The initial velocity
distribution functions are drifting Maxwellians with temperatures
$T_e=1/12$ and $T_i=5/12$ for electrons and ions, respectively; both
temperatures are initially uniform over the whole domain.  We use
these temperature values so that $v_{th,e}$ and $v_{th,i}$ are similar
and a common velocity space bin size can be used (see
Appendix~\ref{appendix:binning}). The density is set to balance plasma
pressure in the fluid sense, with $n(y)={\rm sech}^2 [(y-L_y/4)/w_0] +
{\rm sech}^2 [(y-3L_y/4)/w_0] +n_b$, where $n_b=0.2$ is the background
(lobe) density.  Therefore, the total upstream plasma $\beta$ for
  this simulation is $n_b k_B (T_e + T_i) / (B_0^2 / 2 \mu_0) = 0.2$.
Unlike the Geospace Environmental Modeling (GEM) magnetic reconnection
challenge simulations \citep{birn:2001}, there is only one Maxwellian
component in the current sheet.  The ion-to-electron mass ratio
$m_i/m_e=25$ and the speed of light $c$ is 15.  These choices enforce
that the plasma is non-relativistic (the speed of light exceeds the
thermal and Alfv\'en speeds), which is appropriate for the
non-relativistic treatment of kinetic entropy being considered here.

We use a small enough spatial grid scale and time step to ensure
excellent conservation of energy and minimize numerical dissipation.
We employ a time step of $\Delta t=0.001 \ \Omega_{ci}^{-1}=0.025
\ \Omega_{ce}^{-1}=0.075 \ \omega_{pe}^{-1}$, which is a factor of
about 2.67 smaller than what would typically be used for these simulation
parameters.  The smallest electron Debye length for this simulation
(based on the maximum density of $1 + n_b$) is $\lambda_{De}=0.018$.
We select a grid scale of $\Delta x=\Delta y=0.0125\approx 0.6944
\ \lambda_{De}$, again smaller than what is typically used for these
simulation parameters to improve energy conservation.

Additional to the parameters for the PIC simulation, the 
kinetic entropy diagnostic requires a number of other parameters, 
which are discussed in detail in Appendix~\ref{appendix:implement}.  
These parameters are only for the kinetic
entropy diagnostic; they do not influence the rest of the simulation.
As discussed in Appendix~\ref{appendix:macrovsreal}, in order to calculate the
combinatorial Boltzmann entropy $\mathcal{S}$ properly, the number of
actual particles per macro-particle $a$ has to be specified at run
time.

We first estimate $a$ using the method described in
Appendix~\ref{appendix:macrovsreal}. For the ``base'' simulation, the particle
weight is proportional to the local density at $t = 0$, with a value
of $W = 0.2/1.44$ in the lobe and $W=1.2/1.44$ at the center of
current sheet.  We use $PPG = 100$ in the base simulation and, as
calculated above, a grid scale of $\Delta x = 0.6944
\ \lambda_{De}$. To relate to the actual number of particles, we
appeal to the system of interest being simulated.  For a simulation
representing the plasma in a solar active region, Table~1 gives $n
\lambda_{De}^3 \simeq 1.3 \times 10^7$, so Eq.~(\ref{eq:ncell1}) gives
$N_{cell} \simeq 4.3 \times 10^6$ actual particles per grid cell.
Using $W = 0.2/1.44$, Eq.~(\ref{eq:ncella}) gives $a = 3.13 \times
10^5$ actual particles per macro-particle.  For the plasma sheet in
Earth's magnetotail, Table~1 gives $n \lambda_{De}^3 \simeq 1.0 \times
10^{13}$, so assuming the same weight and grid scale gives $a = 2.5
\times 10^{11}$.  For what we refer to as the ``base'' simulation, we
use $a=3.13\times 10^5$.

We also need to choose the velocity space bin size $\Delta v$ and the
initial number of macro-particles per grid cell $PPG$ per species.
For each, we must optimize these parameters, which is discussed in
detail in Secs.~\ref{sec:a_dependence} - \ref{sec:deltav_dependence}.
For the base simulation, we use $\Delta v = 1$ and $PPG = 100$. The velocity range for binning the particles is from $-12$ to $12$ in each dimension. Since the plasma is in the non-relativistic regime in this simulation, the choice of a broader velocity range than this should not make much difference.

\section{Results}
\label{sec:results}

The layout of this section is as follows.  We start with a validation
of the implementation of the kinetic entropy diagnostics in the code
in Sec.~\ref{sec:entropycalc}.  The time evolution and conversion of
energy and kinetic entropy is discussed in
Sec.~\ref{sec:conservation}. We discuss the position and velocity
space entropies in Sec.~\ref{sec:posvel}.
Sections~\ref{sec:a_dependence}-\ref{sec:deltav_dependence} contain
results on varying $a$, $PPG$, and $\Delta v$, respectively.  Unless
otherwise noted, the results presented here employ the implementation
discussed in Appendix~\ref{appendix:implement} on the base simulation described
in Sec.~\ref{sec:simulations}.

\subsection{Validation of the Kinetic Entropy Diagnostic}
\label{sec:entropycalc}

\begin{figure}
  \includegraphics[width=3.4in]{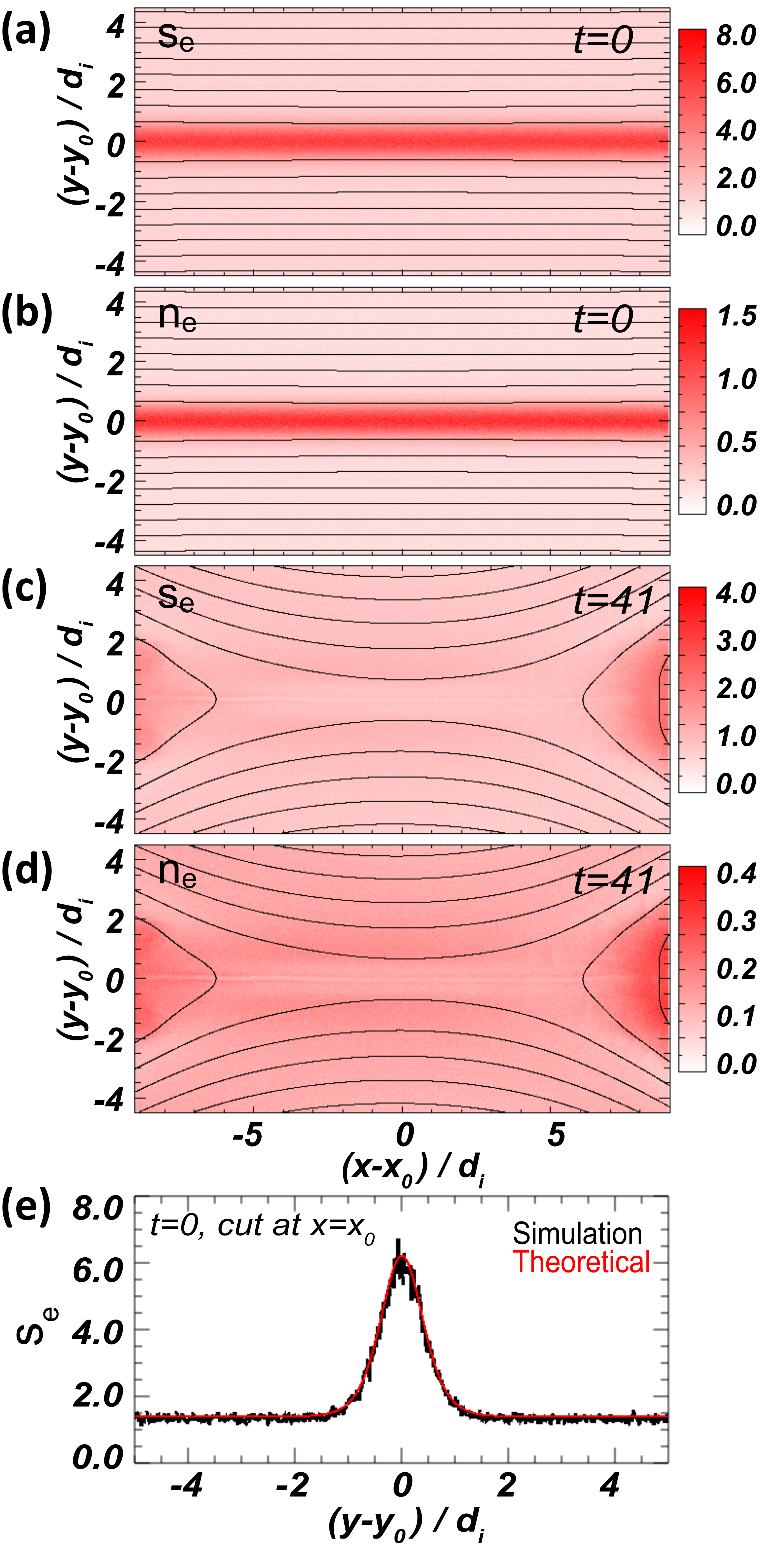} \centering
%  \plotone{Fig1_flnf_ne_t0_t4_v2.pdf}
\caption{2D plots, zoomed in near the reconnection X-line at
  $(x_0,y_0)$ of (a) electron kinetic entropy density $s_e$ and (b)
  electron density $n_e$ at time $t = 0$.  (c) and (d) are the same
  except at $t = 41$.  (e) A vertical cut of $s_e$ through the X-line
  (black) at $t = 0$, with the theoretical prediction (red)
  overplotted.}
  	\label{fig1:entropy_verify}
\end{figure}
Fig.~\ref{fig1:entropy_verify}(a) shows a 2D plot of the continuous
Boltzmann entropy density $s_e(\vec{r})$ from
Eq.~(\ref{eq:entropy_density_sec2}) at time $t = 0$ for electrons; results
for ions are analogous.  The center of the plot is shifted to the
position of the X-line $(x_0,y_0)$ of the top current sheet at $y_0 =
3L_y/4$.  Panel (b) shows the electron density $n_e$ at the same time.
The structure of $s_e$ is strongly determined by the density, as
expected from Eq.~(\ref{eq:entropy_density_maxwell_sec2}) for Maxwellian
distributions such as those at the initial conditions of the present
simulations.  Panels (c) and (d) show similar plots, but for $t = 41$,
showing a similar relationship between kinetic entropy and density
even though distribution functions are no longer all Maxwellian at
this time.

The initial distribution functions for this simulation are drifting
Maxwellians, so we can validate the implementation of the diagnostic
by comparing the calculated $s_e$ with the analytic calculation in
Eq.~(\ref{eq:entropy_density_maxwell_sec2}).  In the upstream region where
the density is 0.2, Eq.~(\ref{eq:entropy_density_maxwell_sec2}) predicts a
value (in normalized code units) of
$(3/2)(0.2)[1+\ln(2\pi(1/12)/(0.04 \times 0.2^{2/3}))] = 1.39$; in the center
of the sheet where the density is 1.2 the analytic prediction is 6.21.
Panel (e) shows a vertical cut of the continuous Boltzmann entropy
density at $t = 0$ in black, with the analytical prediction
overplotted as the red line, revealing excellent agreement of the
theory and simulations.  In Sec.~\ref{sec:a_dependence}, we confirm
that the combinatorial $\mathcal{S}$ and continuous $S$ Boltzmann
entropies are in agreement, as they should be.  We conclude that the
kinetic entropy diagnostics implemented here successfully determine
the kinetic entropy.

\subsection{Energy and Kinetic Entropy Conservation and Conversion}
\label{sec:conservation}

A principal diagnostic of momentum-conserving PIC codes is the
conservation of total (particle plus electromagnetic) energy.
Departures from perfect conservation occur only as a result of
numerical effects including finite time step, finite grid scale, and
noise introduced by having a finite number of macro-particles.  In a
collisionless PIC code, as is the case for the one employed in this
study, kinetic entropy should also be conserved \citep{Bellan08}, with
departures from perfect conservation again only arising due to
numerical effects.  Here, we investigate energy and kinetic entropy
conservation in our base simulation.

\begin{figure}
\includegraphics[width=3.4in]{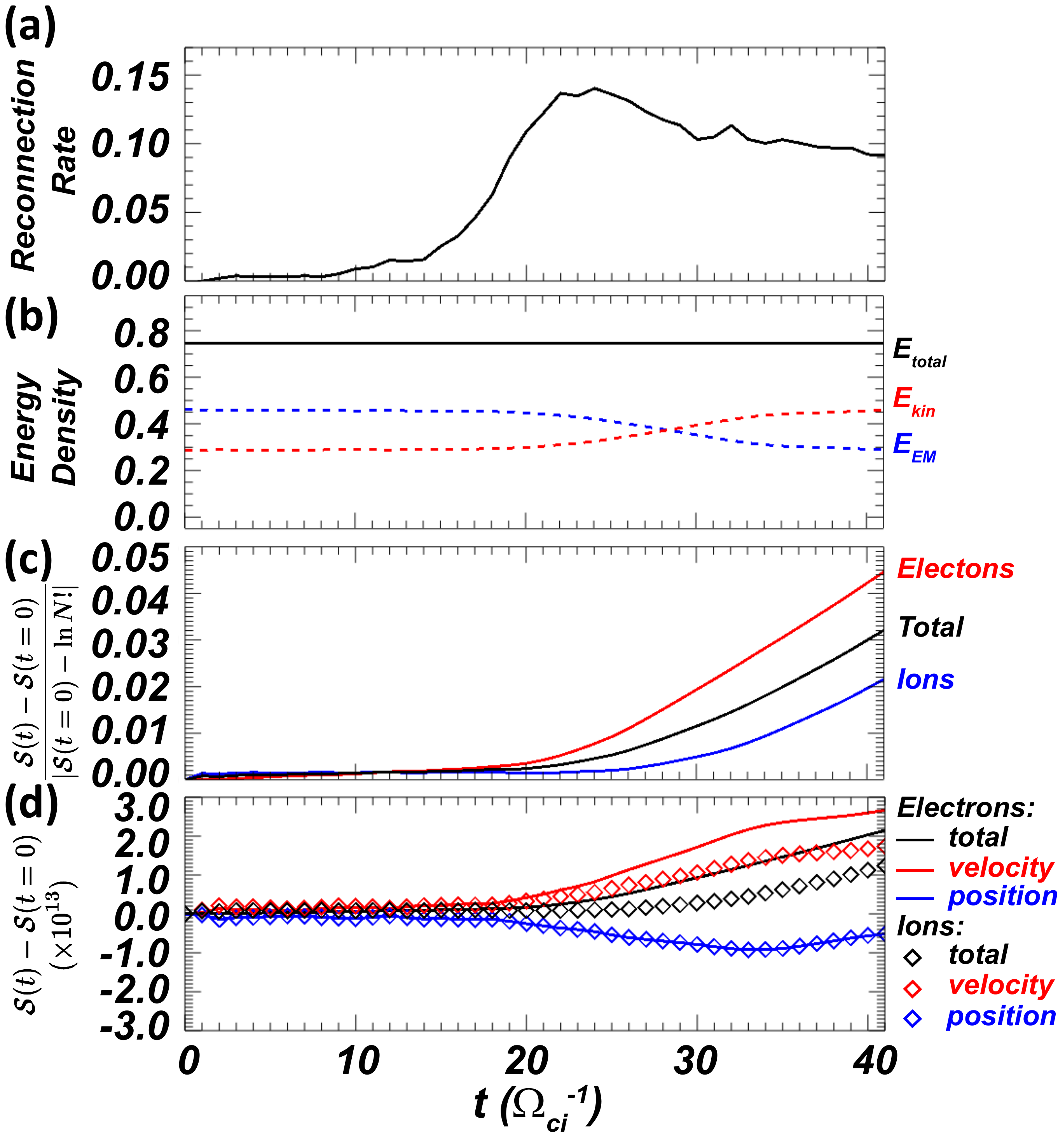} \centering
% \plotone{Fig2_time_history_v3.pdf}
\caption{Time histories from the base simulation of the following
  quantities: (a) reconnection rate, (b) total energy density
  $E_{total}$ (black solid line), total kinetic energy density
  $E_{kin}$ (red dashed line), and electromagnetic energy density
  $E_{EM}$ (blue dashed line), (c) relative change of the non-constant
  term in the combinatorial Boltzmann entropy $\mathcal{S}(t)$ in
  Eq.~(\ref{eq:lngamma1_sec2}) for electrons (red), ions (blue), and total
  (black), (d) deviation from its initial value of velocity space
  entropy $\mathcal{S}_{{\rm velocity}}$ (red), position space entropy
  $\mathcal{S}_{{\rm position}}$ (blue), and total combinatorial
  Boltzmann $\mathcal{S}$ (black) for electrons (solid curves) and
  ions (diamonds).}  \label{fig2:timehistory}
\end{figure}

The time evolution of the system is shown using the reconnection rate
as a function of time $t$ in Fig.~\ref{fig2:timehistory}(a).  The
reconnection rate is the time rate of change of magnetic flux between
the X-line and O-line, identified at each time $t$ using the saddle
and extremum of the magnetic flux function $\psi(\vec{r})$ defined by
$\vec{B} = \hat{z} \times \nabla \psi$, where $\vec{B}$ is the
magnetic field.  As is typical in 2D PIC simulations in periodic
domains, the reconnection rate starts to grow from zero (visibly at $t
\approx 10$), reaches a peak (at $t \approx 21.5$), and then falls
back down to a reasonably steady state (for $t > 34$).

Fig.~\ref{fig2:timehistory}(b) shows total energy density
$E_{total}$ (black solid curve), total kinetic energy density
$E_{kin}$ (red dashed curve) including both bulk and thermal kinetic
contributions, and total electromagnetic energy density $E_{EM}$ (blue
dashed curve), as a function of time $t$ for the base simulation.  The
total energy only increases 0.24\% by $t = 41$; this is excellent
total energy conservation.  This is the result of our intentional use
of a small time step and grid scale.  The expected conversion of
electromagnetic energy to kinetic energy during the reconnection
process (starting in earnest at about $t = 20$) is also seen in the
time histories.

Now, we investigate how the kinetic entropy changes in time during the
simulation, including both relative and absolute changes in kinetic
entropy since both provide useful insights.  For the relative change
of the combinatorial Boltzmann entropy $\mathcal{S}$ in
Eq.~(\ref{eq:lngamma1_sec2}), it is important to note that
$[\mathcal{S}(t) - \mathcal{S}(t=0)] / \mathcal{S}(t=0)$ is not a
meaningful measure of the relative kinetic entropy change .  This
  is because the combinatorial Boltzmann entropy $\mathcal{S}$ can be
  written as a sum of two terms [see Eq.~(\ref{eq:lngamma1_sec2})],
and the first term is a large constant term. Thus, calculating
  the relative change in kinetic entropy merely as $[\mathcal{S}(t) -
    \mathcal{S}(t=0)] / \mathcal{S}(t=0)$ would be misleading, because
  each has a large term that does not change but skews the ratio. For
  this reason, we subtract out the constant term and report the change
  in kinetic entropy relative to the initial portion of the
  combinatorial kinetic entropy that can change, which is
  $\mathcal{S}(t=0) - k_B \ln N!$.

Fig.~\ref{fig2:timehistory}(c) shows the change of the combinatorial Boltzmann entropy in
time from Eq.~(\ref{eq:lngamma1_sec2}) normalized to $\mathcal{S}(t=0)-k_B
\ln N!$ for the base simulation, with values for electrons in red,
ions in blue, and their total in black.  The relative changes are
about 4.5\%, 2.1\% and 3.2\% by $t$=41 for electrons, ions, and total,
respectively. In general, the kinetic entropies are conserved
reasonably well, given that reconnection occurs and there is a
conversion of nearly one-third of the electromagnetic energy into
particle kinetic energy.  Interestingly, the kinetic entropy due to
numerical effects is monotonically increasing.  If the code had
physical collisions, one would expect the kinetic entropy would
monotonically increase.  We find that the numerical effects, in this
sense, mimic physical collisions.

The absolute change to the kinetic entropy is now used to study the
partition between electrons and ions.
Fig.~\ref{fig2:timehistory}(d) shows the total combinatorial
Boltzmann entropy $\mathcal{S}$ (in black) for electrons (solid line)
and ions (diamonds) as a function of time $t$.  Each has its initial
value subtracted so that the plotted values are the change relative to
the initial time.  Notice the change in the absolute kinetic entropies
are quite large, at the $10^{13}$ level in code units (corresponding
to the $10^{-10}$ level in units of J/K).  This ostensibly large
number is a result of the number of actual particles represented in
the simulation being large.  In particular, the base simulation has
100 $PPG$ and 4096 $\times$ 2048 cells, for a total of 838,860,800
macro-particles.  With $a = 3.13 \times 10^5$, the total number of
particles represented is $N = 2.6 \times 10^{14}$.  The first term in
the kinetic entropy in Eq.~(\ref{eq:stirling}) is $\ln N!$, which is
approximately $8.5 \times 10^{15}$.  This sets the scale of kinetic
entropies for this system; we find the total kinetic entropies after
the subtraction due to the second term in Eq.~(\ref{eq:stirling}) are
at the $10^{14}$ level, and the change in kinetic entropy in time is
at the $10^{13}$ level, as seen in Fig.~\ref{fig2:timehistory}(d).

Comparing the total kinetic entropies for each individual species, we
see that both increase in time as might be expected, but the electrons
gain more than the ions in an absolute sense.  This is very
reasonable, as numerical effects arising at small scales are expected
to disproportionately affect electrons.

\subsection{Position and Velocity Space Entropies}
\label{sec:posvel}

We now discuss the position and velocity space entropies discussed in
Appendix~\ref{appendix:pos_vel}.  The two terms are calculated from
Eqs.~(\ref{eq:ent_space_sec2}) and (\ref{eq:ent_vel_sec2}) using the
combinatorial Boltzmann entropy $\mathcal{S}$.  Their evolution is
shown in Fig.~\ref{fig2:timehistory}(d), with position space entropies
in blue and velocity space entropies in red with electrons given by
the solid lines and ions by the diamonds.  First, we note that the
position space entropy is essentially the same for electrons and ions.
This is consistent with expectations as a result of quasi-neutrality
of the plasma.

The velocity space entropy increases for both electrons and ions, a
result of a temperature increase of both species due to the
reconnection process, as expected from Appendix~\ref{appendix:pos_vel}.  The
increase in velocity space entropy is associated with a decrease in
the position space entropy.  If kinetic entropy is perfectly
conserved, as the governing equations would have in this closed
system, then any increase in velocity space entropy would necessarily
be offset by a decrease in position space entropy. In the simulation,
total kinetic entropy is not conserved perfectly, but we still observe
a decrease in position space entropy for both electrons and ions.
Physically, this decrease is associated with the enhanced density in
the island as reconnection proceeds and upstream plasma is compressed.
Compression leads to more particles in some phase space bins, lowering
the position space entropy as discussed in Appendix~\ref{appendix:pos_vel}.

This explanation is predicated on the notion that the temperature
increase is physical rather than numerical, so we investigate this
here. The increase of total entropy due to numerical effects is less
than 5\%, as discussed in Sec.~\ref{sec:conservation}.  One might
expect the thermal energy change from numerical effects $\Delta
E_{th,numerical}$ to scale like $Q_{numerical} \simeq T \Delta
S_{numerical}$ from the first law of thermodynamics, so $\Delta
E_{th,numerical}$ would be at the 5\% level. However, in the
simulation, the thermal energy gain for electrons and ions are 103\%
and 77\%, respectively. This implies that physical heating is much
more significant than the contribution due to numerical effects.

This result also underscores a point about temperature and entropy
that is important to take into account in laboratory and satellite
measurements of kinetic entropy.  In this simulation, there is a
significant increase in thermal energy, but only a small change in
kinetic entropy.  This shows that a temperature increase is not
necessarily associated with an increase in total kinetic entropy.

\begin{figure}
 \includegraphics[width=3.4in]{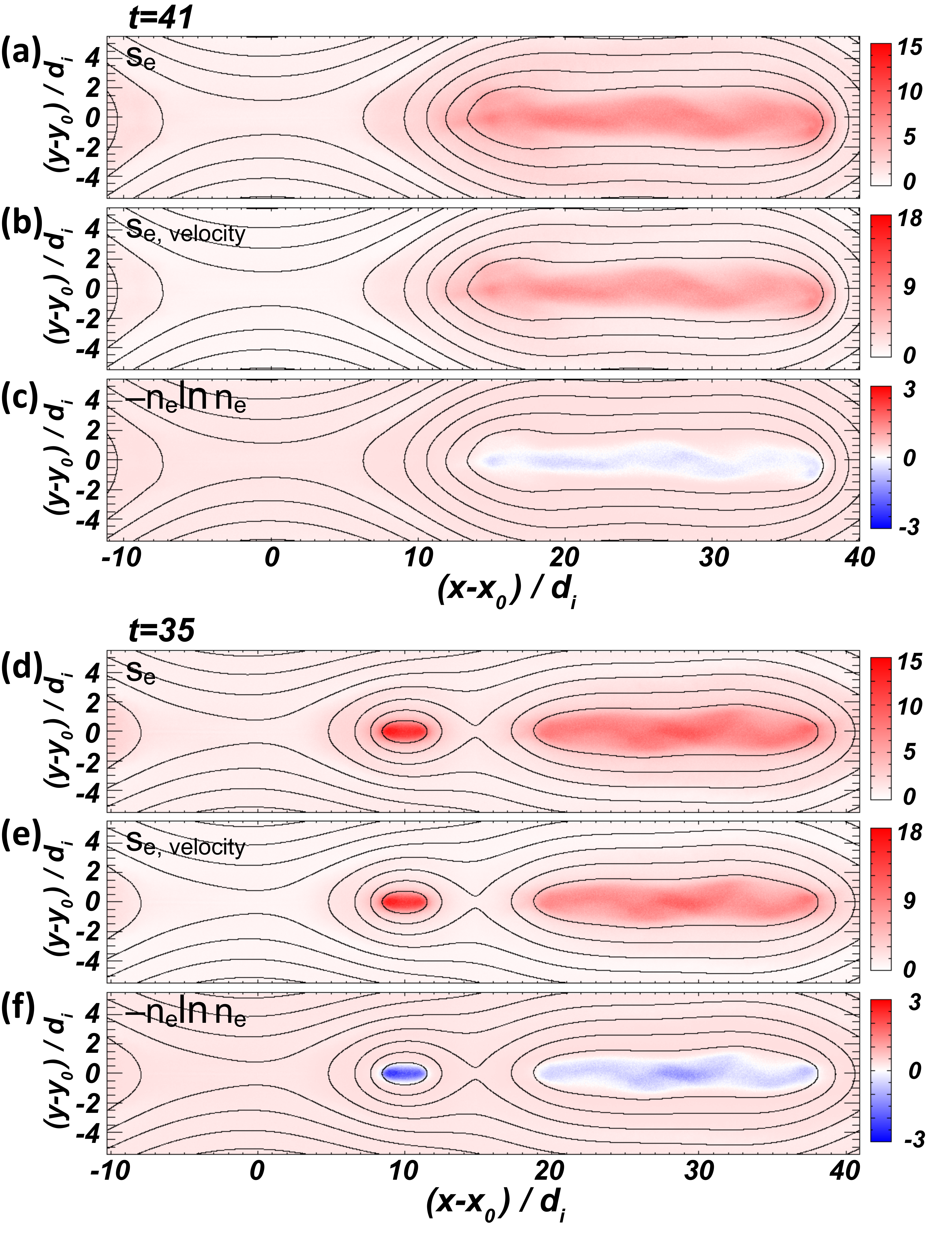}
  % \plotone{Fig3_s_position_velocity_t41_t35_upperpart.pdf}
  \caption{2D plots of various electron kinetic entropies: (a)
      continuous Boltzmann entropy density $s_e(\vec{r})$, (b)
      velocity space entropy density $s_{e,{\rm velocity}}(\vec{r})$,
      (c) the $-n_e(\vec{r}) \ln n_e(\vec{r})$ term that arises in the
      calculation of $S_{\text{position}}$, all evaluated at $t = 41$.
      (d) - (f) are analogous plots at $t = 35$, near the minimum in
      total position space kinetic entropy when there is a secondary
      island further compressing the
      plasma.}  \label{fig3:entropy_space_vel}
\end{figure}

To get a sense for what the different kinetic entropies look like as a
function of space, Fig.~\ref{fig3:entropy_space_vel} includes plots of
(a) continuous Boltzmann entropy density $s_e(\vec{r})$ [from
  Eq.~(\ref{eq:entropy_density_sec2})], (b) velocity space
  entropy density $s_{e,{\rm velocity}}(\vec{r})$ [from
  Eq.~(\ref{eq:svelocity_sec2})], and (c) the $-n_e(\vec{r}) \ln
  n_e(\vec{r})$ density related to the position space entropy [from
  Eq.~(\ref{eq:space_stirling_lim_sec2})], each evaluated at $t =
  41$.  These plots are all for electrons and are showing the whole
domain in $x$ and are zoomed in to the upper current sheet in $y$. 

Caution is needed in interpreting these plots.  The regions of
  highest entropy in panels (a) and (b) do not necessarily reflect
  regions of increased kinetic entropy because the kinetic entropy at
  $t=0$ is not uniform in space since the plasma density is higher
  close to the center of the initial current sheet, as is shown near
  the current sheet in Fig.~\ref{fig1:entropy_verify}(a).
Similarly, assessing the temporal change in total kinetic
  entropy, as plotted in Fig.~\ref{fig2:timehistory} is
  non-trivial solely from these plots, because
Fig.~\ref{fig2:timehistory} represents the total kinetic entropy
  integrated over all space.  Thus, assessing the change in total
  kinetic entropy at later times requires integrating the 2D plots in
Fig.~\ref{fig3:entropy_space_vel} over all space and comparing
  with the initial integrated kinetic entropy.

Panels (a) and (b) reveal elevated levels of kinetic entropy in
  the islands, which is the combined result of the higher density
  (higher entropy) plasma initially in the current sheet getting
  corralled into the island, and the plasma in the island being heated
  which increases its velocity space kinetic entropy.  The blue swath
  in the island in panel (c) shows that the change in the position
  space kinetic entropy is negative there, which is consistent with
  the plasma being compressed in the island.

Further evidence of this interpretation is shown in
Figs.~\ref{fig3:entropy_space_vel}(d)-(f) which has plots
  analogous to panels (a) - (c) but evaluated at $t = 35$, near the
  global minimum in position space kinetic entropy as seen in
Fig.~\ref{fig2:timehistory}(d). There is clearly a secondary
  island clearly present near $(x-x_0,y-y_0) = (10,0)$, and the island
  has a significant decrease of $-n_e(\vec{r}) \ln n_e(\vec{r})$ where
  compression is most significant.  This justifies the stated comment
  that compression in the islands leads to a decrease in position
  space kinetic entropy.  For the parameters in the base simulation,
the difference between $s_e(\vec{r})$ and $s_{e,{\rm
    velocity}}(\vec{r})$ is at about the 10\% level.

\subsection{Importance of Including Actual Particles Per Macro-particle
  for the Combinatorial Boltzmann Entropy}
\label{sec:a_dependence}

As discussed in Appendix~\ref{appendix:macrovsreal}, to calculate the
combinatorial Boltzmann entropy $\mathcal{S}$, one must include the
number of actual particles per macro-particle $a$.  Here, we show this
is the case in the simulations.  Furthermore, since the combinatorial
$\mathcal{S}$ and continuous $S$ Boltzmann entropies should be nearly
identical for a large number of particles, and the two are coded
  in separately rather than $S$ following from $\mathcal{S}$ from the
  explicit use of Stirling's approximation, we can use this as a
further test of the implementation of the diagnostics.

\begin{figure}
\includegraphics[width=3.4in]{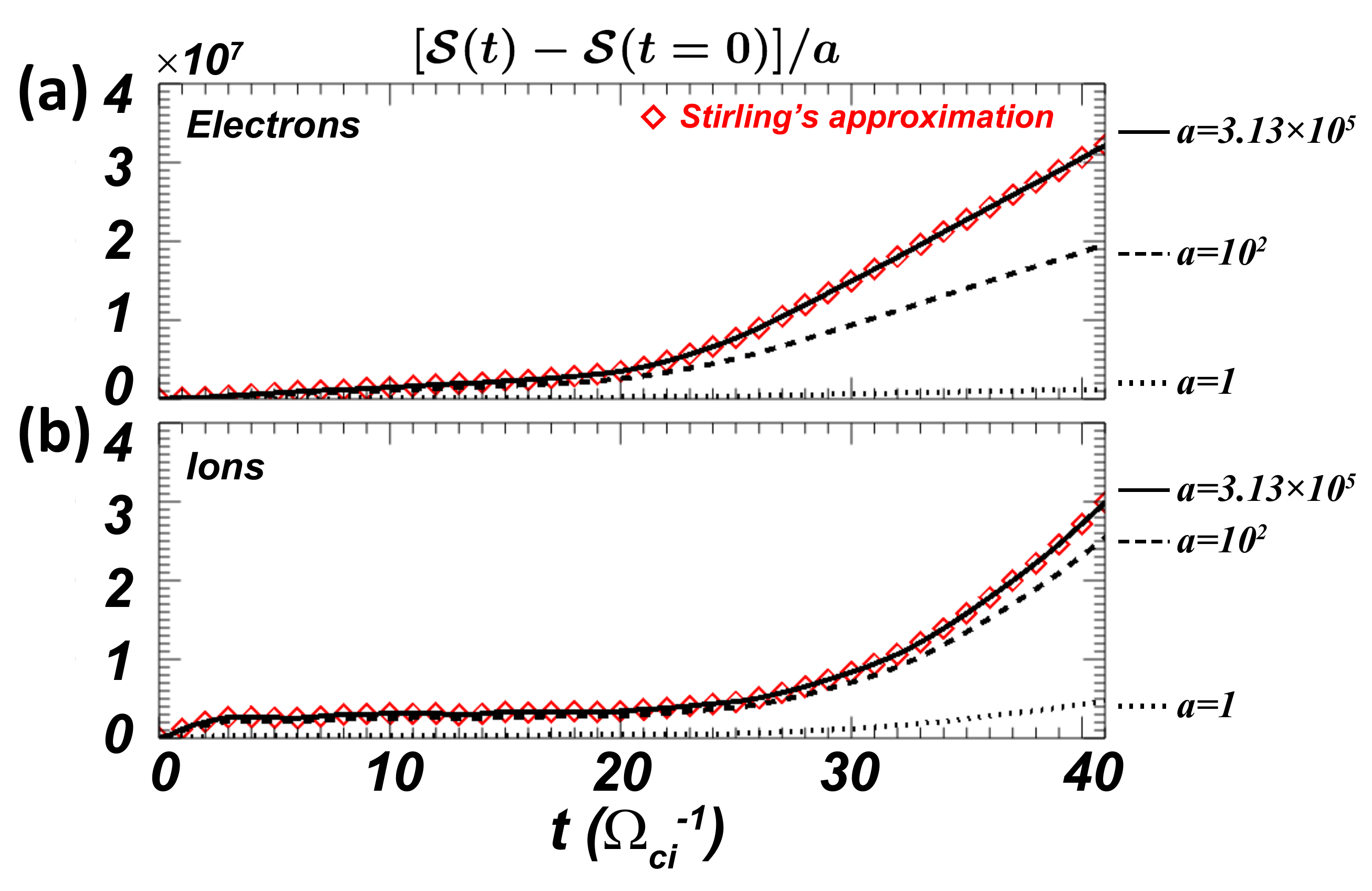}
% \plotone{Fig4_compare_a1_a2_a5_norm_updated_v2.pdf}
\caption{Combinatorial Boltzmann entropy deviations from their initial
  value normalized to $a$, {\it i.e.,}
  $[\mathcal{S}(t)-\mathcal{S}(t=0)]/a$ for (a) electrons and (b)
  ions.  Solid, dashed, and dotted lines are for $a =3.13\times 10^5,
  10^2$ and 1, respectively.  The red diamond symbols indicate the
  value for the continuous Boltzmann entropy $S$ from
  Eq.~(\ref{eq:stirling_f2_sec2}).} \label{fig4:dependencea}
\end{figure}

We perform three simulations that are identical except for the use of
different values of $a$.  An $a = 1$ case has each macro-particle
representing a single particle, and we also use values of $a = 100$
and the base simulation using $a = 3.13 \times 10^5$.  The $a=1$
  case warrants further discussion; one could be concerned that there
  are not enough particles to maintain the plasma
  approximation. However, that is not the case for our simulations.
  Our simulations employ $PPG=100$ for each species.  The (position
  space) grid cell in our simulation is about $(2/3)
  \lambda_{De}$. Thus, these 2D simulations have approximately
  $(3/2)^2 \times 100 = 225$ particles per Debye sphere.  This is much
  larger than 1, as is required for the plasma approximation, and is
  only a factor of two or so lower than the number of particles per
  Debye sphere in the MRX experiment and Earth's ionosphere, as shown
  in Table~\ref{table-plasparam}. Thus, the plasmas being
  simulated continue to satisfy the plasma approximation, even with
  $a=1$.

Fig.~\ref{fig4:dependencea} contains results for the time evolution of
the total combinatorial Boltzmann entropy $\mathcal{S}(t)$ integrated
over the entire computational domain, shown as a difference from its
initial value $S(t=0)$ and divided by $a$, for the three simulations.
Panel (a) is for electrons and panel (b) is for ions.  The reason to
divide by $a$ is that we know from Eq.~(\ref{eq:stirling_a2}) that the
continuous Boltzmann entropy $S$ is directly proportional to $a$ in
the limit of large number of particles, so dividing by $a$ allows us
to directly compare simulations that use different values of $a$.  The
red diamonds show the corresponding value of the kinetic entropy from
Eq.~(\ref{eq:stirling_f2_sec2}), which follows after employing the
Stirling approximation.

First, we note that there is excellent agreement in
the large $a$ simulation between the combinatorial $\mathcal{S}$ and
continuous $S$ Boltzmann entropies as there should be, which provides
additional evidence for the proper implementation of the diagnostic.
For the $a = 100$ case, a significant difference between the two is
observed, especially for the electrons.  For the $a = 1$ case, the
difference is at least an order of magnitude.  The results show that
if $a$ is not included, or is too low, the combinatorial Boltzmann
entropy $\mathcal{S}$ does not agree with the continuous Boltzmann
entropy $S$.

To be more specific, a typical maximum value of macro-particles in a
phase space bin is approximately 3 in the base simulation.  Taking
into account the particle weight of $W = 0.2/1.44$, analogous to the
discussion in Appendix~\ref{appendix:macrovsreal} leading to
Eq.~(\ref{eq:ncell2}), for a simulation with $a = 100$ implies that
there are a maximum of about $3 \times (0.2/1.44) \times 100 \simeq 40$ actual
particles in any phase space bin.  The error due to the Stirling
approximation for an argument of 40 is about 1\%.  While this is
reasonably good, it represents the minimum error in any cell.  Bins
with fewer particles contribute higher errors (4 actual particles has
a 15\% error), leading to the larger errors approaching 30\% we see
for the $a = 100$ simulation.  For $a = 3.13 \times 10^5$, the maximum
particles per cell is 130,000, for which the error introduced by the
Stirling approximation is exceedingly small ($3 \times 10^{-4} \%$).
This motivates the approximate level of disagreement for the $a = 100$
simulation and why the larger $a$ gives good agreement.  We note that
there are a number of physical systems for which $a$ would be of order
100 for $PPG$ near 100 and a weight of $W = 1$, such as Earth's
ionosphere, the MRX reconnection experiment, and high energy density
laser plasmas, as seen in Table~1, so there are physical systems for
which errors could be introduced by using the Stirling approximation.

Fig.~\ref{fig4:dependencea}  indicates that use of the combinatorial form of the kinetic entropy requires the use of the number of real particles per macro-particle $a$ to get physically appropriate results for real systems. In contrast, the continuous $f \ln f$ form of kinetic entropy does not require inclusion of $a$ to get physical appropriate results. (Fig.~\ref{fig4:dependencea} also provides validation that the implementation of the $a$ factor in the PIC code was carried out successfully.) A corollary of this is that it would not be appropriate to run a PIC simulation with the idea that macro-particles represent single particles. Instead, one must take into account the fact that macro-particles represent a large number of real particles in physical plasma systems, or one gets a wrong answer for the combinatorial kinetic entropy. Given that the combinatorial version of the kinetic entropy is a perfectly viable approach to calculate the entropy, it is important to make this point here.

\subsection{Dependence on Macro-particles Per Grid Cell ($PPG$)}
\label{sec:ppg_dependence}

The limited number of macro-particles in PIC simulations leads to a
worse statistical representation of phase space than in the actual
system being simulated.  Here, we investigate how this impacts the
calculation of kinetic entropy by comparing simulations with different
numbers of macro-particles per grid cell, keeping the actual number of
particles fixed by keeping $a$ times $PPG$ constant.  This ensures
there are a sufficient number of particles to avoid accuracy issues as
discussed in Sec.~\ref{sec:a_dependence}.  We carry out simulations
with $PPG$ of 1, 25, 50, and the base simulation of 100.  For $PPG=50,
25,$ and 1, we use $a=6.27\times10^5$, $1.25\times10^6$, and
$3.13\times10^7$, respectively. The reasons we include a case with
$PPG$=1 are (1) some studies have used low $PPG$ in PIC simulations
and (2) we can test what happens to the kinetic entropy calculation
when the statistics are poor.

Some extra details for the $PPG=1$ case are warranted.  Since
numerical PIC noise is expected to be significant, we start by
performing a simulation with the same divergence cleaning frequency as
the other simulations (every 10 particle time steps).  We find the
time history of the reconnection rate is very different than the
higher $PPG$ simulations due to the numerical noise and relatively bad
energy conservation.  Then, we perform another simulation with
divergence cleaning at every time step, which reduces the impact of
the noise.  The total energy change in this simulation is 7.3\%, and
the reconnection rate evolution is similar to the higher $PPG$
simulations.  We find the magnitude of the kinetic entropy change is
similar to the $PPG=1$ case with less frequent divergence cleaning.
Consequently, we use the $PPG = 1$ simulation with the higher cadence
divergence cleaning in what follows.

\begin{figure}
\includegraphics[width=3.4in]{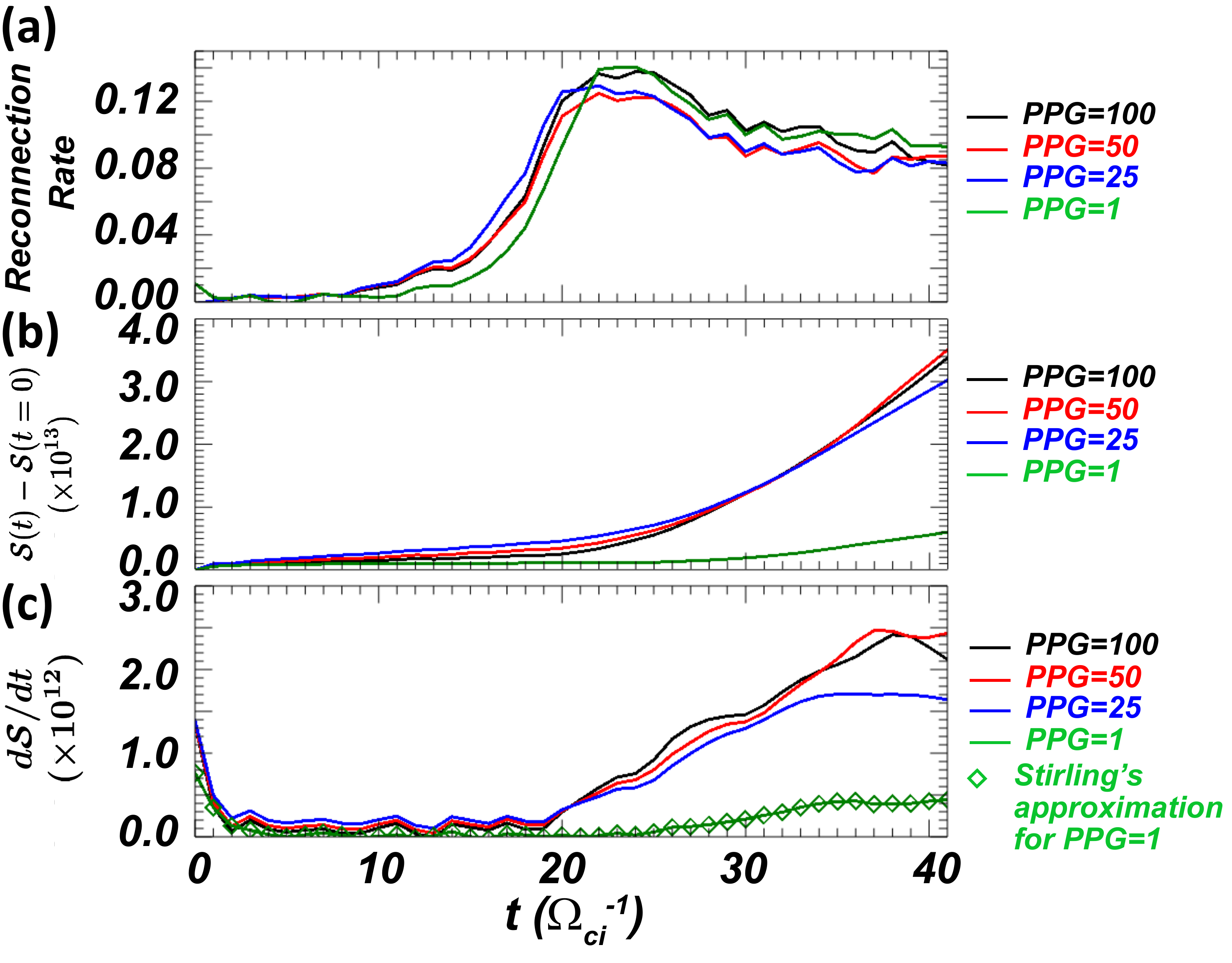}
% \plotone{Fig5_comparePPG_time_history_v3.pdf}
\caption{(a) Reconnection rate, (b) deviation of the total
  combinatorial Boltzmann entropy $\mathcal{S}$ from its initial
  value, and (c) time rate of change of the total combinatorial
  Boltzmann entropy $\mathcal{S}$ for simulations with different $PPG$
  of 100 (black), 50 (red), 25 (blue), and 1 (green). In (c), the
  diamonds show the corresponding value using the continuous Boltzmann
  entropy $S$ instead of the combinatorial Boltzmann entropy
  $\mathcal{S}$ for the $PPG$=1 case to confirm it is calculated
  properly.} \label{fig5:ppg1}
\end{figure}

Fig.~\ref{fig5:ppg1}(a) shows the reconnection rate as a function of
time for the four simulations, with the colors defined in the plot and
caption.  The plot clearly shows that the reconnection rate is quite
insensitive to $PPG$, even for a value of $PPG$ = 1 (with additional
divergence cleaning).  That the reconnection rate can be accurately
simulated in PIC simulations with few particles has been previously
noted in astrophysical PIC simulation studies of reconnection
\citep{Sironi14,Sironi16,Ball18}.

Panel (b) shows the deviation of the combinatorial Boltzmann entropy
$\mathcal{S}$ from its initial value for the four simulations with
different $PPG$.  The $PPG$=1 case deviates from the others
significantly, but the results of the other three cases are
similar. In order to examine the differences among $PPG$=100, 50 and
25, we further plot the time rate of change of the combinatorial
Boltzmann entropy $d\mathcal{S}/dt$ in panel (c). The results for the
$PPG$ = 50 and 100 cases are quite similar.  This suggests that these
numbers for $PPG$ are sufficient to give a relatively stable regime of
the kinetic entropy calculation for our simulations.

In contrast, the $PPG = 25$ results differ from the higher $PPG$
results, showing that adverse numerical effects from the worse
particle statistics take place, especially late in time after
reconnection occurs.  It is even more dramatic for $PPG = 1$, where
there is a large discrepancy approaching an order of magnitude.
Moreover, a 2D plot of the kinetic
entropy density of the $PPG = 1$ simulation (not shown) is very
similar to the density, as expected, but the departure of the
distribution from a Maxwellian has very large noise which swamps out
all other structures (since a Maxwellian is not well described by a
single macro-particle).    These
  important differences suggest that even though a $PPG$ of 1 can be
made to reasonably produce the reconnection rate, one must proceed
with caution on matters related to kinetic entropy, including effects
such as particle acceleration and plasma heating.  A convergence test
of kinetic entropy and the effect of small $PPG$ on energization,
heating, and energy partitioning would be useful in testing such
simulations.

It may seem counter-intuitive that the change of kinetic entropy decreases with
fewer $PPG$ since the simulation should be more noisy when $PPG$
  is low and one might think this would increase the entropy.
  However, there is a subtle reason this is not the case, as we can
  see with an extreme example.  Consider a simulation with only a
  single macro-particle corresponding to $a$ real particles.  All $a$
  real particles corresponding to that macro-particle are in the same
  cell in phase space. The kinetic entropy of this macro-particle is
  equal to that of all $a$ particles in a single cell of phase space
  (which is zero). Now let time evolve. The macro-particle moves to a
  new cell in phase space. Since the macro-particle still corresponds
  to all $a$ particles, all $a$ particles move to the same new cell in
  phase space. Thus, their contribution to the kinetic entropy at this
  later time is exactly the same -- it is still zero. Consequently,
  kinetic entropy is perfectly conserved for this simulation even
  though the number of macro-particles is only 1. Moreover, the low
  number of macro-particles makes the total entropy smaller than it
  would be if there were more $PPG$.  Thus, a decrease in $PPG$
  counter-intuitively leads to a decrease in the change of kinetic entropy despite
  the increase in particle noise. 

\subsection{Dependence on $\Delta v$}
\label{sec:deltav_dependence}

While the kinetic entropy should not depend on grid scale for the
continuous form in Eqs.~(\ref{eq:stirling_f2_sec2}) and~(\ref{eq:stirling_f2}), the discrete form in
Eq.~(\ref{eq:stirling_f}) is required for implementation in PIC and
therefore is dependent on the grid scale.  Here, we discuss how to
choose the size of the velocity space bin size $\Delta v$.  The
dependence on spatial grid size could be determined using the same
approach, but this is left for future work.  We choose the optimal
$\Delta v$ by comparing simulation results for different $\Delta v$ to
analytical results for known Maxwellian distributions at $t=0$ in the
base simulation.

\begin{figure}
\includegraphics[width=3.4in]{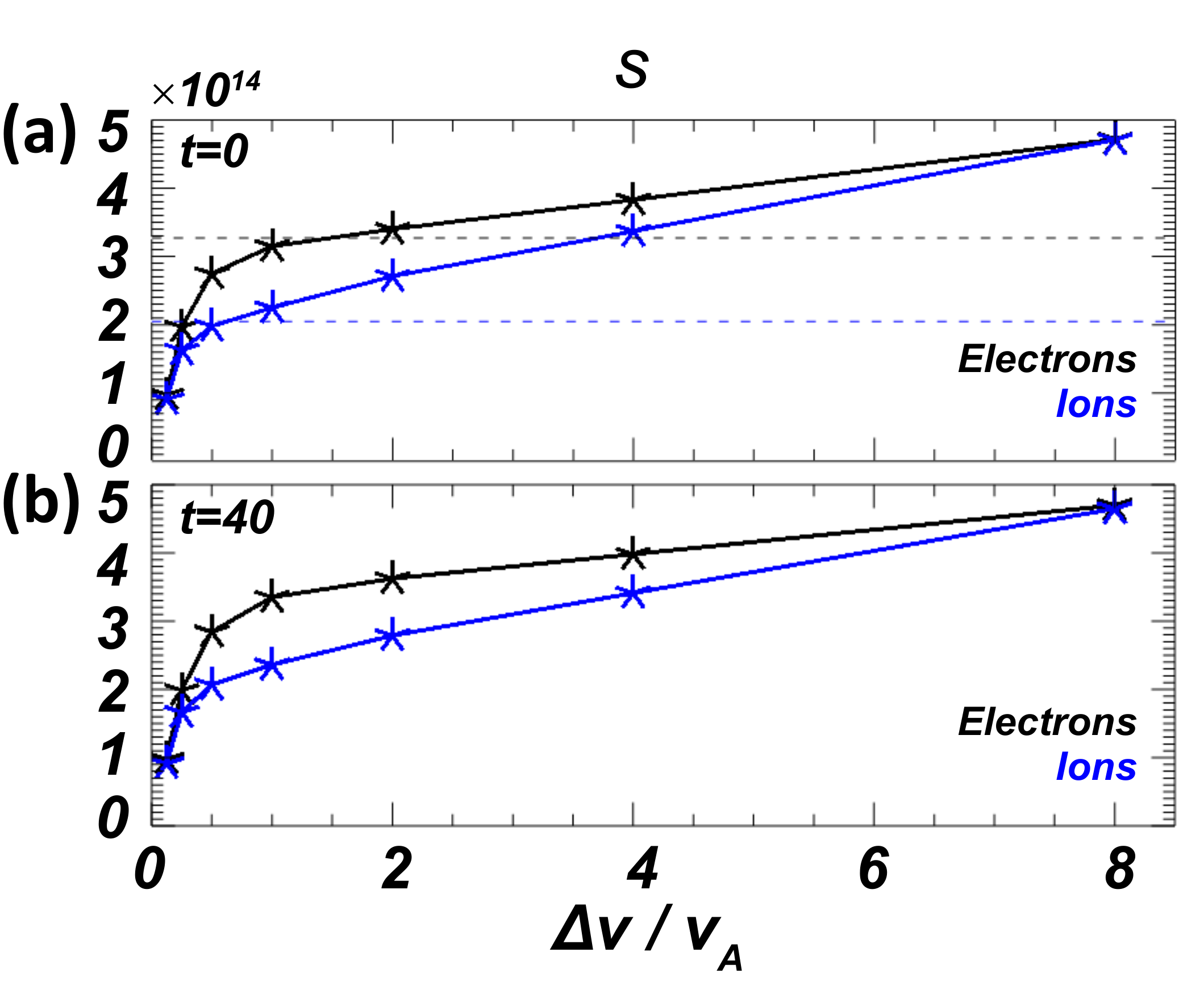} \centering
% \plotone{Fig6_flnf_vs_dv_v3.pdf}
\caption{Continuous Boltzmann entropy $S$ for electrons (black) and
  ions (blue) in seven simulations with $\Delta v/v_A=0.125, 0.25,
  0.5, 1.0, 2.0, 4.0, 8.0$ (a) at $t=0$ and (b) at $t=40$. The dashed
  lines in (a) indicate the analytical values at $t=0$ for electrons
  (black) and ions (blue).} \label{fig6:deltavdependence}
\end{figure}
We show results from multiple simulations using velocity bin sizes
$\Delta v$ of $0.125, 0.25, 0.5, 1.0, 2.0, 4.0,$ and 8.0 relative to
the ion Alfv\'en speed $v_A$.
Fig.~\ref{fig6:deltavdependence}(a) shows the continuous
Boltzmann entropy $S$ at  the initial time $t =
0$ for both electrons (black) and
ions (blue) as a function of the velocity space grid scale $\Delta v$
normalized to the ion Alfv\'en speed $v_A$.  As expected, the
continuous Boltzmann entropy $S$ of both species increases with
$\Delta v$ for sufficiently large values.  Below $\Delta v / v_A$ of
about 0.5 or 1, the variation strongly depends on $\Delta v$.

Also in panel (a) are black and blue horizontal dashed lines
corresponding to the analytical prediction of the continuous Boltzmann
entropy of electrons and ions, respectively, for the initial
conditions from the spatial integral of
Eq.~(\ref{eq:entropy_density_maxwell_sec2}).  By inspection, we see that
the numerically calculated value of electron kinetic entropy agrees
well with the analytical value for a velocity grid scale just over 1
$v_A$.  This suggests an appropriate value to use for the velocity
space grid of electrons.  Similarly, the ion kinetic entropy agrees
with the analytical value for a velocity grid just under 1 $v_A$.
These two results motivated our choice of a grid scale of $\Delta v =
1 v_A$, which is $\approx 0.69 \ v_{th,e}$ in terms of the initial
electron thermal speed $v_{th,e}$ for this simulation.  That this is
slightly less than the electron thermal speed is consistent with
expectations, as discussed in Appendix~\ref{appendix:binning}.  Note, for both
electrons and ions, the velocity grid scale that gives best agreement
with the analytical calculation is near the species thermal speed
($1.44 \ v_A$ for electrons, $0.65 \ v_A$ for ions).  Also, the base
simulation used the same velocity space grid scale for ions and
electrons; this is not a requirement and could be relaxed.

While this approach can be used at $t = 0$ when all the
  distribution functions are Maxwellian and exact solutions are known,
  there is no assurance that the velocity space grid scale will
  continue to be sufficient at later times.  One way to address this
  would be to test systems for which the distribution functions are
  known analytically as a function of time, such as the bump on tail
  instability \citep{Oneil65,Valentini12}. We leave such an approach
  for future work.  More generally, given that phase space evolution
  can lead to very sharp structures in velocity space, this is a very
  fundamental issue that has previously arisen in Vlasov modeling
\citep{Servidio15,Camporeale16,Roytershteyn18},  and it likely has no
  general solution.

That said, we perform further analysis to assess whether the
  velocity space resolution adversely impacts our study at later
  times.  First, we note that the temperature ({\it i.e.}, the spread
  of the distribution function in velocity space) in this reconnecting
  system tends to increase in time throughout the domain, so this
  suggests the resolution at $t=0$ may remain sufficient at later
  times, at least in these simulations.  That this is the case can be
  seen in Fig.~\ref{fig6:deltavdependence}(b), which is
  analogous to panel (a) but at $t = 40$.  The results are quite
  similar to those at $t = 0$, suggesting only a minor global effect.
  Indeed, the global change in kinetic entropy is at the 3\% level for
  this simulation.

A more careful approach is to identify the most non-Maxwellian
  electron distribution in the system at the end of the simulation,
  $t=41$, and test the effect of the velocity space grid scale in
  finding its kinetic entropy density.  For the base simulation, the
  most non-Maxwellian distribution occurs at the X-line at late time,
  when the electrons undergo meandering orbits and produce familiar
  characteristic distributions like those in Fig. 4 of Ng et
al.\citep{Ng11} This distribution function has sharp structure
  and therefore is the hardest to resolve in velocity space, so the
  error of its kinetic entropy density should be the most.

Using this local distribution in a single grid cell, we calculate
  the electron kinetic entropy density as a function of velocity space
  grid scale (not shown), which represents the local counterpart to
  the global result in Fig.~\ref{fig6:deltavdependence}. As in
  the global results, we find that there is a medium range between
  about $0.5v_A$ and $2v_A$ where the entropy is not strongly
  dependent on the velocity space grid.  The uncertainty in the
  kinetic entropy density as a result of the velocity space grid scale
  is approximately 15\%, in spite of the fact that the late time
  distribution function has structures in velocity space that are not
  likely to be completely resolved.

The key point to assess this result is that the change in the
  kinetic entropy between $t$ = 0 and $t$ = 41 is approximately a
  factor of 2, from about 1.3 (for the electrons far upstream of the
  current sheet at $t=0$) to about 0.7 (for the meandering electrons
  at the X-point $t=41$). Thus, the 15\% uncertainty introduced by
  even the worst velocity space grid resolution in our entire
  simulation is considerably smaller than the physical change in
  entropy of nearly a factor of 2.  This shows that the velocity space
  grid scale resolution is sufficient for the purposes of this study.
  However, we emphasize that a careful convergence study is important
  for future studies and in other plasma applications.

\section{Discussion and Conclusion}
\label{sec:conc1}

\subsection{Summary}

This manuscript presents a study of how to implement two forms of the
kinetic entropy into fully kinetic particle-in-cell simulations and
how to use these quantities to diagnose the physical system.  The two
forms are the combinatorial Boltzmann entropy $\mathcal{S} = k_B \ln
\Omega$ and the continuous Boltzmann entropy $S = -k_B \int d^3r d^3v
f \ln f$.  These forms of kinetic entropy, can be decomposed into a
sum of two terms describing the kinetic entropy in position space and
velocity space separately.

We then discuss how to implement the diagnostic into PIC simulations,
including considerations such as the optimal size of the velocity
space grid scale, the number of macro-particles per grid cell, and the
number of actual particles per macro-particle.  We compare and
contrast the merits of each of the two measures of kinetic entropy.

Then, we validate the implementation using two-dimensional in position
space, three-dimensional in velocity space collisionless PIC
simulations of anti-parallel symmetric magnetic reconnection.  The
initial conditions contain only drifting Maxwellian distributions
which has an analytical solution for the kinetic entropy.  This allows
for a careful validation of the implementation at the initial
  time and provides an avenue for optimizing the velocity space grid
size.  Finally, we discuss the interpretation of the results and how
to extract physical understanding from the kinetic entropy.

The results of the present study include the following:
\begin{enumerate}
\item The ``base'' simulation with very low $\Delta t$ demonstrates
  good conservation of the total kinetic entropy (to 3.2\%).  The
  increase in kinetic entropy is purely numerical, but increases
  monotonically as would be expected for physical collisions and
  increases faster when reconnection proceeds.  The level of
    increase of kinetic entropy is small enough that simulations with
    a collision operator should produce entropy at a level high enough
    to be resolved in future studies.
\item Electrons and ions show different kinetic entropy production
  rates, with electrons gaining more than ions in the base simulation
  because their dynamics occurs at smaller scales and therefore are
  disproportionately impacted by numerical effects.
\item We apply the decomposition of kinetic entropy into position
  space and velocity space portions to a numerical system and use it
  to interpret the physics of the system for the first time. Although
  the total kinetic entropy is nearly conserved, the position and
  velocity space entropies $S_{{\rm position}}$ and $S_{{\rm
      velocity}}$ vary noticeably in time.  For both electrons and
  ions, $S_{{\rm position}}$ decreases in time (for most of the
  simulation), while  $S_{{\rm velocity}}$ increases in
  time.  This is physically related to the electrons and ions getting
  heated during reconnection (increasing their velocity space entropy)
  and getting compressed (decreasing their position space
  entropy). This approach will be useful for distinguishing
    reversible and irreversible dissipation in future studies that
    incorporate a collision operator, even for distribution functions
    that are strongly non-Maxwellian.
\item Calculating the combinatorial Boltzmann entropy $\mathcal{S}$
  requires specifying the number of actual particles per
  macro-particle $a$ for the calculation, while the continuous
  Boltzmann entropy only needs this quantity to convert to real units
  for comparison with observations or experiments.
\item We show how to choose  the
  number of macro-particles per grid cell $PPG$.  For these
  simulations, a bin size that is close to the electron thermal speed
  is a good size, and we need at least 50 $PPG$ to get reliable
  kinetic entropy values for our choice of time step and spatial grid
  scale.  The minimum $PPG$ that is sufficient to reliably calculate
  the kinetic entropy likely depends on these quantities.
\item We show how to choose the velocity space bin size at the
  initial time when the simulation has distributions such as
  Maxwellians for which the entropy is attainable analytically.  We
  find a grid scale slightly smaller than the species thermal speed is
  a good bin size for our base simulation. There is no clear path for
  ensuring the velocity space bin size remains adequate for later
  times because sharp velocity space structures are common in weakly
  collisional systems. However, for the present study, we have shown
  that the least resolved distribution at late time introduces only a
  15\% error in our simulation, far smaller than the physical
  difference in the kinetic entropy, so the velocity bin resolution is
  good enough for the purposes of this study.  Future work on this
  issue, for reconnection and for other problems in plasma physics,
  will be very important.
\item We show that the kinetic entropy is not reliably produced
  in simulations with a low number of particles per grid, even though
  the same simulations can be made to reliably produce the
  reconnection rate. This has important implications about studies of
  heating and dissipation in systems with few particles per grid
  cell.
\end{enumerate}
Our study shows that kinetic entropy can serve as a diagnostic of the
fidelity of a collisionless PIC code, alongside the often used energy,
but also can give key physical insights about the dynamics of a
system. The diagnostic developed here should be applicable to any
explicit PIC simulation, which should make it useful in many
heliospheric, planetary, and astrophysical processes including
magnetic reconnection, plasma turbulence, and collisionless shocks.
It is useful for systems with distributions with a thermal core
  and non-thermal tails, but also more broadly for systems with
  strongly non-Maxwellian distributions.

\subsection{Other Insights and Applications}

This work provides a number of other insights that
are important for applying the kinetic entropy diagnostic for
applications.  Kinetic entropy in a PIC simulation is sensitive to the
phase space bin size, both in position and velocity space.  This is
because the calculation is discretized on a finite grid.  Comparisons
between different times in a given simulation, between two different
simulations, and between simulations and data should be done with a
fixed position and velocity space grid scale to the extent possible.

An interesting result is that one needs to be careful to ensure the bins
in phase space have a large number of (actual) particles to obtain
accurate kinetic entropy values.  Stirling's approximation is good to
within 1\% when the number of actual particles in a bin is 40 but has
15\% error for 4 actual particles in a bin.  Thus, computational and
observational studies alike should monitor the number of particles per
phase space bin.  It is possible in either setting to have
insufficient counts to render the Stirling approximation valid.  In
such cases, the combinatorial Boltzmann entropy $\mathcal{S}$ in
Eq.~(\ref{eq:Boltzmann}) is needed over the continuous Boltzmann
entropy $S$ in Eqs.~(\ref{eq:stirling_f2_sec2}) and~(\ref{eq:stirling_f2}).  As discussed in
Sec.~\ref{sec:a_dependence}, this is the case for some important
plasma settings, potentially including laboratory experiments, Earth's
ionosphere, and laser plasmas.

We point out the importance of ensuring a stable regime of the kinetic
entropy with the number of numerical macro-particles per grid cell
$PPG$.  For the base simulation with small time step and well-resolved
grid, we find we need at least 50 for $PPG$ to have a stable regime of
the kinetic entropy.  There have been a number of studies, especially
in the plasma astrophysics community, with smaller $PPG$ including as
low as 1-4 \citep{Sironi14,Sironi16,Ball18}.  We confirm their results
that one can get a reasonable reconnection rate in such systems, but
for our code the low $PPG$ is insufficient to get a proper kinetic
entropy.  The \citet{Ball18} study tested convergence of particle
energy spectra with $PPG$ of 4 and 16; it would be interesting to also
check stability of the kinetic entropy diagnostic.  We suggest that
using kinetic entropy to test for stability for low $PPG$ simulations
is a useful technique which is potentially important for studies of
particle acceleration and plasma heating in reconnection, turbulence,
and shocks.

One challenge for applications is that the conservation of kinetic
entropy in ideal (collisionless) systems is only valid for closed,
isolated systems.  This can easily be accomplished in idealized
simulations, but it is unlikely to be the case in naturally occurring
systems.  The expectation of this line of research is that the
dissipation physics can be studied using idealized simulations, and
then the insights obtained from the simulations can be compared to
real systems.  This is already being carried out with data from MMS
and will be the subject of future publications.

Another challenge is that typically the continuous Boltzmann entropy
density $s = -\int d^3v f \ln f$ is mostly proportional to the number
density, so a plot of kinetic entropy density by itself is unlikely to
reveal any new insights.  We will demonstrate in a follow up study 
that kinetic entropy can be useful for identifying
non-Maxwellian distributions for electrons and ions and furthermore
that the kinetic entropy can be used to estimate the effective
numerical collisionality of a collisionless PIC code.

The initial implementation of the kinetic entropy diagnostic has many
ways to be improved, which we outline here.  First, our treatment is
non-relativistic, but the PIC code in use and many natural systems
relevant to study with this tool are relativistic
\citep{Kaniadakis09}.  In addition, comparisons to implicit PIC
simulations (which can employ much larger spatial grids and time
steps) and Vlasov simulations (which have no PIC noise) would be
interesting.  More in depth studies into the dependence of the kinetic
entropy diagnostic on spatial grid scale and time step would be
useful, along with higher macro-particles per grid cell $PPG$.
Significant work is needed to choose velocity space bin sizes
  that do not introduce larger errors after the initial time.  Our
work used only the linear shape function; it would be interesting to
test other shape functions.  It would also be interesting to examine
kinetic entropy in PIC simulations with open boundary conditions.  The
present simulations are 2D in position space and 3D in velocity space;
simulations that are 3D in both position and velocity space should be
carried out.  Most importantly, this work employs only collisionless
PIC simulations, which means that any dissipation ({\it i.e.,} any
increase of total kinetic entropy) that occurs is through numerical
effects.  Thus, we are unable to address physical mechanisms for
dissipation in the present study.  Using a collisional PIC code would
allow for an investigation of the physical mechanisms of dissipation
with the kinetic entropy diagnostic.

There are also numerous physics topics that are important for future
work.  Future work should also address parametric studies of
  kinetic entropy in magnetic reconnection, as well as in plasma
turbulence and collisionless shocks.  Generalizations to other forms
of entropy, such as the Tsallis entropy which describes long-range
interactions and contains memory effects \citep{Tsallis88}, should
also be undertaken.  Whether chaotic behavior is sufficient to produce
an entropy increase should also be the subject of future work.  It is
important to see if numerical kinetic entropy production can impact
other physical processes like particle acceleration and heating.

\begin{acknowledgments}
We acknowledge helpful conversations with A.~Glocer, H.~Hietala,
W.~Paterson, S.~Schwartz, and E.~G.~Zweibel.  We thank J.~Burch for
motivation for this project.  The authors thank Mahmud Hasan Barbhuiya for comments on the manuscript. Support from NSF Grants AGS-1460037,
AGS-1602769, and PHY-1804428 and NASA Grant NNX16AG76G is gratefully
acknowledged.  S. S. acknowledges the European Union's Horizon 2020 research and innovation programme under grant agreement No 776262 (AIDA, 
www.aida-space.eu); V.R. acknowledges NSF-DOE grant DE-SC0019315; E.E.S. acknowledges NSF grant PHY-1617880; M.A.S. acknowledges NASA grant NNX17AI25G. This research uses resources of the National Energy
Research Scientific Computing Center (NERSC), a DOE Office of Science
User Facility supported by the Office of Science of the
U.S.~Department of Energy under Contract No.~DE-AC02-05CH11231.
\end{acknowledgments}

\appendix
\section{Theory of Kinetic Entropy}
\label{appendix:theory}
In this section, we discuss the theoretical background of kinetic
entropy and its decomposition into position space entropy and velocity
space entropy.

\subsection{Background on Kinetic Entropy}
\label{appendix:background}

For a closed system (which in Nature could be thermally insulated, but
in a simulation can also be periodic), the form of kinetic entropy
$\mathcal{S}$ in a kinetic framework is \citep{Boltzmann77,Planck06}
%\begin{linenomath*}
\begin{equation}\label{eq:Boltzmann}
\mathcal{S}(t)=k_B\ln\Omega(t),
\end{equation}
%%\end{linenomath*}
where $k_B$ is Boltzmann's constant and $\Omega(t)$ is the number of
microstates of the system that produce the system's macrostate at a
time $t$.  In what follows, we suppress the time dependence to
simplify the notation.  Each individual plasma species has its own
associated kinetic entropy, so there is an implicit subscript $e$ or
$i$ for electrons or ions, respectively, that is suppressed for
clarity when possible.  Following the nomenclature in \citet{Frigg11},
we refer to the kinetic entropy in this form as the ``combinatorial
Boltzmann entropy.''  This is one form of kinetic entropy we implement
in our PIC code.

To elucidate the meaning of kinetic entropy in this form, consider a
plasma with a fixed number of charged particles $N$ for each species.
We treat classical, non-relativistic systems (even though the PIC code
we use is fully relativistic).

\begin{figure}
  \includegraphics[width=3.4in]{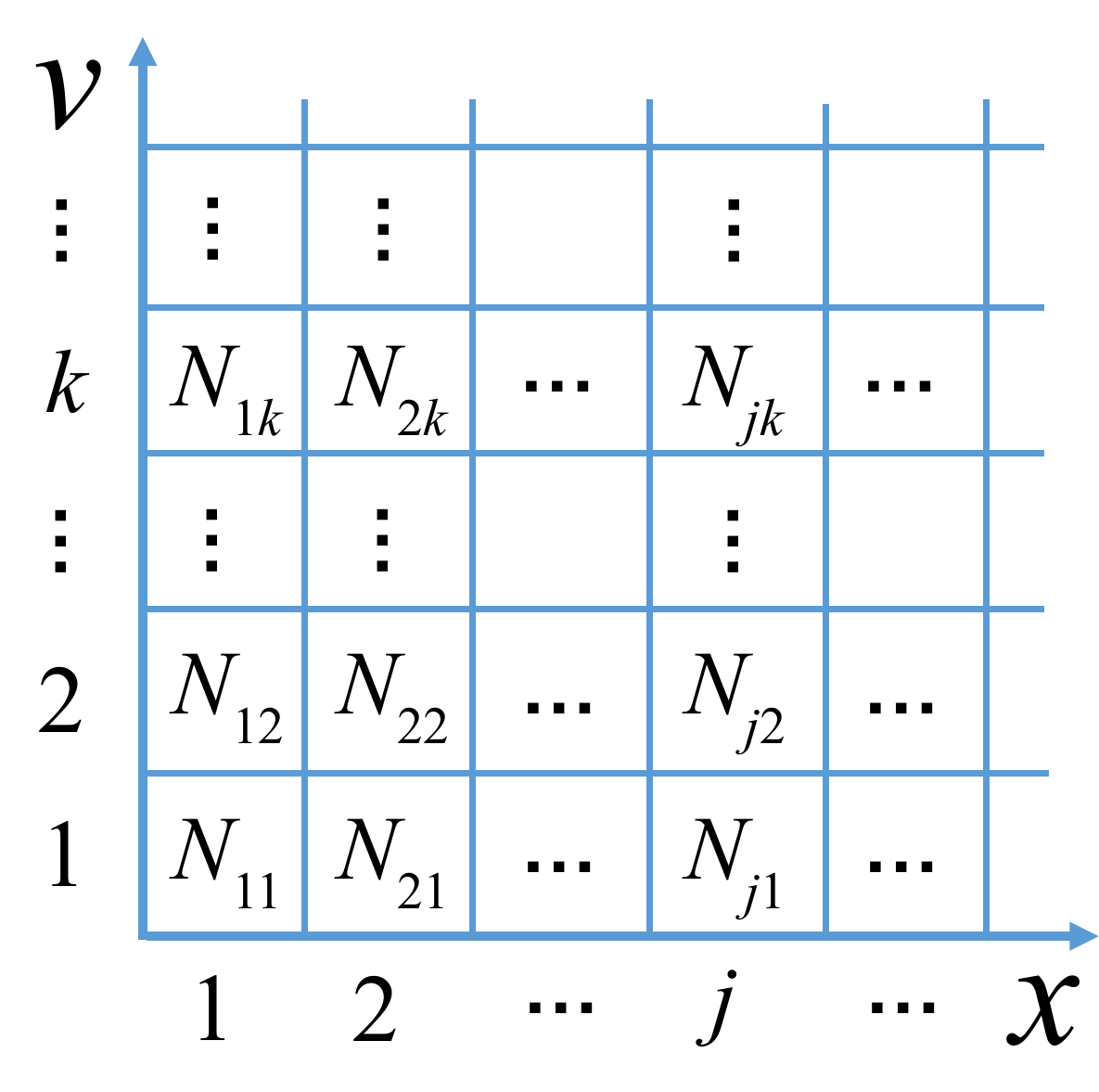}
 % \plotone{Fig1ap_v2.pdf}
  \caption{Sketch of phase space $(x,v)$ for a 1D system, discretized
    into a grid.  The number of particles in the bin spanning position
    $x_j$ to $x_j + \Delta x$ and velocity $v_k$ to $v_k + \Delta v$
    is $N_{jk}$.  This can be suitably extended to higher dimensional
    systems.}
  \label{fig1ap}
  \end{figure}
For a three-dimensional (3D) system, phase space is 6D with each
particle described by its position and velocity $(\vec{r},\vec{v})$.
To calculate kinetic entropy, phase space is discretized into domains
we call bins.  Fig.~\ref{fig1ap} shows the discretization of an
analogous 1D system.  Define $N_{jk}$ as the number of particles in
the phase space bin spanning positions $\vec{r}_j$ to $\vec{r}_j +
\Delta \vec{r}$ and velocities $\vec{v}_k$ to $\vec{v}_k + \Delta
\vec{v}$ at a given time $t$, where the components of $\Delta \vec{r}$
and $\Delta \vec{v}$ describe the extent of the bin in each direction
in phase space.  At this point, we nominally take these bins as finite
in size ({\it i.e.,} not infinitesimal) with an eye to calculating
kinetic entropy in PIC simulations.  The volumes of the bins in
position and velocity space are $\Delta^3r$ and $\Delta^3v$,
respectively.  In a 1D system, subscripts $j$ and $k$ signify the bin
in position space and velocity space, respectively.  In 3D, we
continue to use $j$ and $k$ as shorthand to identify the bin, even
though we actually need to specify each component of the position and
velocity to identify a bin.  Thus, we think of $j$ to mean
$j_x,j_y,j_z$ for the $x$,$y$,$z$ directions in position space and $k$
to mean $k_x,k_y,k_z$ for the $v_x$,$v_y$,$v_z$ directions in velocity
space.  By definition,
%\begin{linenomath*}
\begin{equation}
N = \sum_{j,k}N_{jk}. \label{eq:nsum}
\end{equation}
%%\end{linenomath*}
A given macrostate is defined by the collection of all the $N_{jk}$,
which via integration yields all the fluid quantities of the system.
A microstate is a possible way to choose the particles in the system
to produce a given macrostate, treating individual particles
classically as distinguishable.

Using this construct, the number $\Omega$ of possible microstates for
a given macrostate is calculated using combinatorics \citep{Bellan08}; 
it is the number of permutations that produce the
macrostate with $N_{jk}$ particles in the $jk$th cell by swapping
individual distinguishable particles between any of the bins, {\it
  i.e.},
%\begin{linenomath*}
\begin{equation}
  \Omega=\frac{N!}{\prod_{j,k}N_{jk}!}.
\end{equation}
%%\end{linenomath*}
Inserting this expression into Eq.~(\ref{eq:Boltzmann}) and
simplifying gives the combinatorial Boltzmann entropy $\mathcal{S}$ in terms
of the $N_{jk}$:
%\begin{linenomath*}
\begin{equation}
\mathcal{S}=k_B\left[\ln N!-\sum_{j,k}\ln N_{jk}!\right]. \label{eq:lngamma1}
\end{equation}
%%\end{linenomath*}
The first term is a constant assuming the total number of particles
$N$ in the closed system is fixed.  Since only changes in entropy are
physically important, we can drop the first term if desired (though we
retain it in the calculation of the combinatorial Boltzmann entropy in
our PIC simulations).  Note, however, that whether the first term is
retained or not, quantities like percentage changes in entropy should
be calculated solely relative to the second term.

It is common to approximate Eq.~(\ref{eq:lngamma1}) using Stirling's
approximation $\ln N_{jk}! \approx N_{jk}\ln N_{jk}-N_{jk}$, which is
valid when $N_{jk} \gg 1$, as is typically the case but may have
exceptions.  A short calculation using Eq.~(\ref{eq:nsum}) yields
%\begin{linenomath*}
\begin{equation}\label{eq:stirling}
S = k_B\left[N\ln N-\sum_{j,k}N_{jk}\ln N_{jk}\right],
\end{equation}
%%\end{linenomath*}
where we write the approximate entropy as $S$ instead of
$\mathcal{S}$.  For use in a kinetic description of a fluid or plasma,
one writes the kinetic entropy in terms of the distribution function
$f(\vec{r},\vec{v})$.  The distribution function at position
$\vec{r}_j$ and velocity $\vec{v}_k$ is approximated as
%  \begin{linenomath*}
  \begin{equation}
    f(\vec{r}_j,\vec{v}_k) \approx \frac{N_{jk}}{\Delta^3r\Delta^3v}.
    \label{eq:def_f}
\end{equation}
%%\end{linenomath*}
Replacing $N_{jk}$ in Eq.~(\ref{eq:stirling}) with this expression and
simplifying gives
%\begin{linenomath*}
\begin{eqnarray}
S = & k_B & \left[ N \ln\left(\frac{N}{\Delta^3r \Delta^3v}\right)
  \right. \nonumber \\ & & \left. - \sum_{j,k} (\Delta^3r \Delta^3v)
  f(\vec{r}_j,\vec{v}_k) \left[\ln f(\vec{r}_j,\vec{v}_k) \right]
  \right]. \label{eq:stirling_f}
\end{eqnarray}
%%\end{linenomath*}
As in Eq.~(\ref{eq:lngamma1}), the first term is a constant (for a
fixed phase space bin size) and can be discarded.  In the limit in
which $\Delta\vec{r}$ and $\Delta\vec{v}$ are small, the second term
yields the commonly used form of the kinetic entropy
%\begin{linenomath*}
\begin{equation}\label{eq:stirling_f2}
S = - k_B \int d^3{r}d^3v f(\vec{r},\vec{v}) \left[\ln
  f(\vec{r},\vec{v}) \right],
\end{equation}
%%\end{linenomath*}
where $d^3r$ and $d^3v$ are the infinitesimal spatial and velocity
space volumes.  Following the nomenclature of \citet{Frigg11}, we
refer to Eq.~(\ref{eq:stirling_f2}) as the ``continuous Boltzmann
entropy'' to distinguish it from the combinatorial Boltzmann entropy
$\mathcal{S}$. This is the second form of kinetic entropy we implement
in our PIC code. Note that in dropping the first term of
Eq.~(\ref{eq:stirling_f}), there is an issue with the units of $S$ in
that the second term is no longer formally dimensionless.  Therefore,
care is necessary when the continuous Boltzmann entropy is desired in
proper units.  We discuss this in more detail in
Appendix~\ref{appendix:calcrealunits}.

We note in passing that one can alternately normalize $f$ to be a
probability density rather than a phase space density.  In this
convention, the entropy would be related to the Shannon entropy and
information theory \citep{Shannon48,Jaynes62}.  We do not employ this
convention here with an eye to experiments and observations that
directly measure distribution functions.

The continuous Boltzmann entropy density, {\it i.e.,} the continuous
Boltzmann entropy per unit volume, is denoted by $s(\vec{r})$ and
given by
%\begin{linenomath*}
\begin{equation}
s(\vec{r}) = - k_B \int d^3v f(\vec{r},\vec{v}) \left[\ln
  f(\vec{r},\vec{v}) \right]. \label{eq:entropy_density}
\end{equation}
%%\end{linenomath*}
We point out that the continuous Boltzmann entropy density
$s_M(\vec{r})$ for a 3D drifting Maxwellian distribution in local
thermodynamic equilibrium (LTE) for a species of mass $m$, number
density $n(\vec{r})$, bulk flow velocity $\vec{u}(\vec{r})$, and
temperature $T(\vec{r})$, with $f(\vec{r},\vec{v}) = f_{M} =
n(\vec{r})[m/2 \pi k_B T(\vec{r})]^{3/2} e^{-m [\vec{v} -
    \vec{u}(\vec{r})]^2 / 2 k_B T(\vec{r})}$, is exactly solvable with
%\begin{linenomath*}
\begin{equation}\label{eq:entropy_density_maxwell}
s_M(\vec{r}) = \frac{3}{2}k_B n(\vec{r}) \left[ 1+\ln \left(\frac{2\pi
    k_B T(\vec{r})}{m n^{2/3}(\vec{r})}\right) \right]. 
\end{equation}
%\end{linenomath*}
This result shows the fluid entropy per particle $s/n$ is related to
$p / \rho^\gamma$, where $p = n k_B T$ is the (scalar) pressure, $\rho
= m n$ is the mass density, and $\gamma = 5/3$ is the ratio of
specific heats.  In an adiabatic process, conservation of $s/n$ is
synonymous with conservation of $p/\rho^\gamma$, which is typically
used in fluid models.  Equation~(\ref{eq:entropy_density_maxwell}) is
useful for validating the implementation of the kinetic entropy
diagnostic into kinetic codes.

\subsection{Decomposition of Kinetic Entropy into Position and Velocity
  Space Entropies}
\label{appendix:pos_vel}

Boltzmann's kinetic entropy is defined in terms of permutations of
particles with any position and velocity in phase space.  It is
tempting to interpret the kinetic entropy density in
Eq.~(\ref{eq:entropy_density}) as the entropy purely associated with
permuting particles in velocity space, but this is only correct if the
plasma density is uniform.  If the density is non-uniform ({\it i.e.,}
$n$ is a function of $\vec{r}$), it has been shown that the total
kinetic entropy can be decomposed into a sum of a position space
entropy and a velocity space entropy \citep{Mouhot11,Goldstein04},
as we now review.

By adding and subtracting a common term in Eq.~(\ref{eq:lngamma1}),
$k_B \sum_{j} \ln N_j!$, where $N_j=\sum_{k}N_{jk}$ is the total number of
particles in spatial cell $j$, {\it i.e.,} with any velocity, the
combinatorial Boltzmann entropy $\mathcal{S}$ can be written as
%\begin{linenomath*}
\begin{eqnarray}
\mathcal{S} & = & k_B\left[\ln N!-\sum_{j}\ln N_j! \right] \nonumber
\\ & + & k_B\sum_{j}\left[\ln N_j!-\sum_{k}\ln
  N_{jk}!\right]. \label{eq:space_and_v}
\end{eqnarray}
%\end{linenomath*}
The first two terms have the same form as Eq.~(\ref{eq:lngamma1}),
except that the second term has $N_j!$ instead of $N_{jk}!$, so they
are defined as the position space kinetic entropy,
%\begin{linenomath*}
\begin{equation}\label{eq:ent_space}
\mathcal{S}_{\text{position}}=k_B\left[\ln N!-\sum_{j}\ln N_j!\right].
\end{equation}
%\end{linenomath*}
Similarly, the last two terms in Eq.~(\ref{eq:space_and_v}) have the
same form as Eq.~(\ref{eq:lngamma1}) with $N$ replaced by $N_j$ and
the summation being only over velocity space, so they are defined as
the velocity space kinetic entropy
%\begin{linenomath*}
\begin{equation}
\mathcal{S}_{\text{velocity}} = \sum_j k_B\left[\ln
  N_j!-\displaystyle\sum_{k}\ln N_{jk}!\right]. \label{eq:ent_vel}
\end{equation}
%\end{linenomath*}
Consequently, Eq.~(\ref{eq:space_and_v}) can be written as
%\begin{linenomath*}
\begin{equation}\label{eq:space_plus_v}
\mathcal{S} =
\mathcal{S}_{\text{position}}+\mathcal{S}_{\text{velocity}},
\end{equation}
%\end{linenomath*}
so the combinatorial Boltzmann entropy is decomposed into a sum of
position space kinetic entropy and velocity space kinetic entropy.

Note that there is an asymmetry between the treatment of position and
velocity space in this definition of the position space entropy and
velocity space entropy. The number of microstates per macrostate is
calculated in velocity space for each spatial cell to obtain velocity
space entropy, while the position space entropy is obtained by summing
over velocity space first.  Alternatively, one could interchange the
treatment of position and velocity space in this calculation.
Therefore, the decomposition used here is not unique.  However, the
decomposition employed here and elsewhere gives meaningful information
about local velocity space entropy changes that are indicative of
heating or dissipation, which makes it a preferred decomposition.

As in Appendix~\ref{appendix:background}, one can readily derive expressions
for the position and velocity space kinetic entropies in terms of the
distribution function and analogous expressions in terms of the plasma
density $n$; using Stirling's approximation assuming there are a large
number of particles, one obtains the discrete forms of the continuous
Boltzmann position and velocity space kinetic entropies as
%\begin{linenomath*}
\begin{eqnarray}
S_{\text{position}} & = &
k_B\left[N\ln\left(\frac{N}{\Delta^3r}\right) \right. \nonumber \\ & -
  & \left. \sum_j (\Delta^3r) n(\vec{r}_j)\ln
  n(\vec{r}_j)\right], \label{eq:space_stirling}
%\\ \text{}_{\text{}} & \text{} & \nonumber
%\int d^3r s_{\text{position}}(\vec{r}), 
%\\ \text{}_{\text{}}\text{} & \text{} & \nonumber
%S_{\text{position}} & = & \sum_j (\Delta^3r)
%s_{\text{position}}(\vec{r}_j), \\ s_{\text{position}}(\vec{r}_j) & =
%& k_B\left[n(\vec{r}_j)\ln\left(\frac{N}{\Delta^3r}\right) \right.
%  \nonumber \\ & - & \left.  n(\vec{r}_j)\ln
%  n(\vec{r}_j)\right], \label{eq:space_density_stirling} \\ 
\\S_{\text
  {velocity}} & \equiv & \sum_j (\Delta^3r)
s_{\text{velocity}}(\vec{r}_j), \\ s_{\text{velocity}}(\vec{r}_j) & =
& k_B\left[n(\vec{r}_j)\ln\left(\frac{n(\vec{r}_j)}{\Delta^3v}\right)
  \right.  \nonumber \\ & - & \left.
  \sum_k(\Delta^3v)f(\vec{r}_j,\vec{v}_k)\ln
  f(\vec{r}_j,\vec{v}_k)\right], \label{eq:v_stirling}
\end{eqnarray}
%\end{linenomath*}
where $n(\vec{r}_j)=N_j / \Delta^3r$ is the number density at spatial
cell $j$.  Expressions in terms of continuous variables come from
taking the limit of small bin size gives
%\begin{linenomath*}
\begin{eqnarray}
S_{\text{position}} & = &
k_B\left[N\ln\left(\frac{N}{\Delta^3r}\right) \right. \nonumber \\ &
  -& \left. \int d^3r n(\vec{r})\ln
  n(\vec{r})\right], \label{eq:space_stirling_lim}
%\\ \text{}_{\text{}} & \text{} & \nonumber
%\int d^3r s_{\text{position}}(\vec{r}), 
%\\ \text{}_{\text{}}\text{} & \text{} & \nonumber
%k_B\left[n(\vec{r})\ln\left(\frac{N}{\Delta^3r}\right) \right.
%  \nonumber \\ & -& \left. n(\vec{r})\ln
%  n(\vec{r})\right], \label{eq:sposition} 
\\  S_{\text{velocity}} & \equiv &
\int d^3r s_{\text{velocity}}(\vec{r}),
\\ s_{\text{velocity}}(\vec{r}) & = &
k_B\left[n(\vec{r})\ln\left(\frac{n(\vec{r})}{\Delta^3v}\right)
  \right.  \nonumber \\ & -& \left. \int d^3v f(\vec{r},\vec{v})\ln
  f(\vec{r},\vec{v})\right]. \label{eq:svelocity}
\end{eqnarray}
%\end{linenomath*}

Note, the second term in $s_{\text{velocity}}(\vec{r})$ is merely
$s(\vec{r})$ from Eq.~(\ref{eq:entropy_density}), so the two differ by
the first term.  The key point is that the kinetic entropy density
$-k_B \int d^3v f \ln f$ is not the velocity space entropy because of
this extra term.  Only in the limit in which $n(\vec{r})$ is uniform
are the two effectively the same. .

The physical meaning of the position and velocity space entropies are
given by analogies with the combinatorial Boltzmann entropy
$\mathcal{S}$.  The position space entropy describes the entropy
arising from permutations of particles in position space without
regard to their velocity.  For example, there is only one way to have
all the particles in a single bin in position space; $\Omega = 1$ for
that system and the position space entropy is zero.  In contrast, a
uniform density has the largest number of microstates that produce
that macrostate, so it is the configuration with the largest position
space entropy.  Therefore, compressing a plasma increases the local
density, so is associated with a local decrease in position space
entropy.

The velocity space entropy has a similar interpretation -- it is the
entropy associated with the permutation of particles in velocity space
at a fixed cell in phase space, then summed over all spatial bins.  As
with the position space entropy, more distributed particles in
velocity space are associated with higher velocity space entropy,
while sharper (colder) distributions have lower velocity space
entropies.  Increases in density and temperature both lead to an
increase in velocity space entropy, as is seen explicitly for a
Maxwellian distribution in Eq.~(\ref{eq:entropy_density_maxwell}).
Note, for an adiabatic process for a system in local thermodynamic
equilibrium, the total entropy is conserved.  However, the position
and velocity space entropies can change, with kinetic entropy
converted between them.  During adiabatic compression, for example,
the position space entropy decreases as described above.  This
decrease is perfectly balanced by adiabatic heating which increases
the velocity space entropy.  We find the decomposition into position
and velocity space entropies provides useful insights in the analysis
of the PIC simulations.

\section{Implementation of Kinetic Entropy Diagnostic in PIC Simulations}
\label{appendix:implement}

\begin{table*}
\caption{\label{table-plasparam} Representative values of the plasma parameter $n
    \lambda_{De}^3$ in a number of plasma
    settings \citep{Ji11,Rosenberg15,NRL18}.}
\begin{ruledtabular}
\begin{tabular}{lccc}
 %&\multicolumn{2}{c}{$D_{4h}^1$}&\multicolumn{2}{c}{$D_{4h}^5$}\\
Setting & Density $({\rm cm}^{-3})$ & $T_e$ (eV) & $n \lambda_{De}^3$\\ \hline
Solar active region & $10^9$ & 100 & $1.3 \times 10^7$\\
Magnetotail & 0.2 & 500 & $1.0 \times 10^{13}$ \\
MRX reconnection experiment & (0.1 - 1) $\times 10^{14}$ & 5-15 & 450-7,000 \\
Solar wind at 1 AU & 10 & 10 & $4.1 \times 10^9$ \\
Magnetosheath & 20 & 50 & $3.2 \times 10^{10}$ \\
Earth's ionosphere & $10^6$ & 0.01-0.1 & 410-13,000 \\
High energy density laser plasma & $10^{20}$ & 1000 & 1,300\\
\end{tabular}
\end{ruledtabular}
\end{table*}

In this section, we provide a detailed summary of how we implement the
kinetic entropy diagnostic into our PIC code {\sc p3d}
\citep{Zeiler02}, although the approach should be applicable to any
explicit PIC code.  We emphasize that we use periodic boundary
conditions so that the system is closed and one can unambiguously
determine if there are global changes in kinetic entropy (as opposed
to open systems where the kinetic entropy can change via dynamics at
the boundary).  In what follows, we break down the procedure into
steps and discuss each in turn.

\subsection{Macro-particles vs.~Actual Particles}
\label{appendix:macrovsreal}

As discussed in Appendix~\ref{appendix:background}, calculating the
combinatorial $\mathcal{S}$ or continuous $S$ Boltzmann entropies
requires a knowledge of the number of particles in each cell in phase
space.  In a PIC simulation, the ``particles'' are actually
macro-particles, each representing a chunk of phase space containing a
large number of actual particles.  Therefore, there is a difference
between the number of particles and number of macro-particles in each
cell.  As we show here, the relative structure of the continuous
Boltzmann entropy $S$ is not sensitive to this difference.  However,
when converting $S$ from a PIC simulation into real units, the results
are sensitive to this difference.  Moreover, the combinatorial
Boltzmann entropy $\mathcal{S}$ is sensitive to the number of actual
particles represented by each macro-particle.

Here, we discuss how to relate the number of macro-particles to the
number of actual particles.  We define a constant $a$ as the number of
actual particles per macro-particle.  The approach to estimate $a$ is
to find the number of actual particles, say, electrons, that would be
in a given grid cell in the simulation.  For a system with a known
number density $n$, the number of electrons $N_{cell}$ in a spatial
volume $\Delta^3r$ corresponding to a grid cell in PIC is
%\begin{linenomath*}
  \begin{equation}
    N_{cell} \sim n \Delta^3r. \label{eq:ncell1}
\end{equation}
%\end{linenomath*}
A typical grid size for an explicit PIC simulation is close to the
electron Debye length $\lambda_{De} = (\epsilon_0 k_B T_e / n_e
e^2)^{1/2}$.  Thus, $N_{cell}$ is on a similar scale as the plasma
parameter $n \lambda_{De}^3$.  For reference, representative values
for the plasma parameter in various settings are provided in Table~1,
though of course these are merely representative and may differ for
particular applications.

To get a comparable number for the PIC code in order to find $a$, we
note that many PIC codes, including the one in use here, allow for
macro-particles to be assigned a different weight $W$, which improves
the statistics in systems with non-uniform initial densities.  This
must be accounted for in the estimation of $N_{cell}$.  We now
estimate $N_{cell}$ using the initial conditions of the simulations
carried out for the present study. At $t=0$ in our simulations, $W$ is
same for all macro-particles in each grid cell and is proportional to
the local density.  Thus, $PPG \times W$ represents the effective
number of macro-particles per grid cell, so at $t=0$ the number of
actual particles in a cell is
%\begin{linenomath*}
  \begin{equation}
    N_{cell} = PPG \times W \times a. \label{eq:ncell2}
\end{equation}
%\end{linenomath*}
Equating the two expressions for $N_{cell}$ from
Eqs.~(\ref{eq:ncell1}) and (\ref{eq:ncell2}) gives
%\begin{linenomath*}
  \begin{equation}
    a = \frac{n \Delta^3r}{PPG \times W}. \label{eq:ncella}
  \end{equation}
%\end{linenomath*}
In simulations for which $W$ is not a constant for all particles in
each cell, a generalization of this approach is necessary.

It is important to note when and how including $a$ is necessary in
calculating kinetic entropy.  Define $\mathcal{N}_{jk}$ as the number
of weighted macro-particles in the $jk$th bin in phase space; then
%\begin{linenomath*}
\begin{equation}
  N_{jk} = a \mathcal{N}_{jk}.
\end{equation}
%\end{linenomath*}
The value for $\mathcal{N}_{jk}$ is what one gets from the code when
counting weighted macro-particles, but does not take into account the
number of actual particles per macro-particle. Physically, because the
limited number of macro-particles in a PIC simulation implies that
there is a small number of macro-particles per phase space bin, the
number of permutations of the macro-particles is much smaller than the
number of permutations of actual particles. Therefore, if one uses
$\mathcal{N}_{jk}$ instead of $N_{jk}$ to calculate
Eq.~(\ref{eq:lngamma1}), the result is much smaller than that of
actual system. More importantly, the Stirling approximation and thus
the continuous Boltzmann entropy $S$ definition would be invalid since
$\mathcal{N}_{jk}$ is small. The importance of including $a$ can be
seen analytically, as well.  Writing Eq.~(\ref{eq:lngamma1}) in terms
of $\mathcal{N}_{jk}$ gives $\mathcal{S} = k_B [\ln (a\mathcal{N})!  -
  \sum_{j,k} \ln (a\mathcal{N}_{jk})!]$, which is not equal to $a k_B [\ln
  \mathcal{N}! - \sum_{j,k} \ln \mathcal{N}_{jk}!]$.  Thus, the value for
$a$ must be included at calculation time to get the proper value of
the combinatorial Boltzmann entropy $\mathcal{S}$.

In contrast, the kinetic entropy ({\it i.e.,} after using the Stirling
approximation) is simply linear in $a$.  Using $N = a \mathcal{N}$ and
$N_{jk} = a \mathcal{N}_{jk}$ in Eq.~(\ref{eq:stirling}) gives
%\begin{linenomath*}
\begin{equation}\label{eq:stirling_a}
S = k_B\left[a\mathcal{N}
  \ln(a\mathcal{N})-\sum_{j,k}a\mathcal{N}_{jk}\ln
  (a\mathcal{N}_{jk})\right].
\end{equation}
%\end{linenomath*}
Carrying out simple manipulations gives
%\begin{linenomath*}
  \begin{equation}\label{eq:stirling_a2}
    S = a k_B\left[\mathcal{N}\ln
  \mathcal{N} - \sum_{j,k} \mathcal{N}_{jk}\ln
  \mathcal{N}_{jk}\right].
\end{equation}
%\end{linenomath*}
Thus, one can simply calculate the continuous Boltzmann entropy using
macro-particles in the simulation, and then scale the result by $a$ to
get a value for $S$.  The same result holds for the forms in terms of
the distribution function $f$ [{\it i.e.,} Eq.~(\ref{eq:stirling_f2})
  and (\ref{eq:entropy_density})].  In other words, if comparing $f$
or $S$ between a PIC simulation and observations or experiments and an
absolute comparison is desired, one must multiply the raw $f$ and $S$
from the simulation by $a$ to convert it to a physical result.

\subsection{Binning Macro-Particles in Phase Space}
\label{appendix:binning}

In order to obtain the distribution function, one has to discretize
phase space (with bins from $\vec{r}_j,\vec{v}_k$ to $\vec{r}_j +
\Delta\vec{r},\vec{v}_k + \Delta\vec{v}$) and calculate the
contribution of each macro-particle to every phase space bin.  There
are numerous approaches to representing the number density of a
macro-particle in a PIC code, referred to as its shape
\citep{Birdsall04}.  The approach used in {\sc p3d}, and therefore
applied here, is a linear shaping function that assumes the charge
density from each macro-particle drops linearly from its maximum to
zero a distance one spatial grid cell away in each direction.
Therefore, in any PIC simulation without a $\delta$-function shaping
function, a macro-particle contributes to the density in each of the
surrounding cells.  To calculate kinetic entropy, we use the same
shape function for each particle in velocity space ({\it i.e.,}
linear). Therefore, the number of macro-particles in a phase space bin
at any given time is typically not an integer.  We suggest that the
implementation of the kinetic entropy calculation should employ the
same particle shape as what is employed in the code in use, but leave
further investigation to future work.

Here is the procedure we use for determining the number of
macro-particles in each phase space bin:
\begin{itemize}
\item Without using the kinetic entropy diagnostic, optimize the
  numerical parameters on a test simulation to ensure proper spatial
  and temporal resolution.  Using the output from this simulation,
  find the maximum speed $v_{max}$ among all macro-particles for all
  times, which should be $\ll c$ in the non-relativistic limit.  Then,
  the range of velocity space to be discretized is restricted to
  $[-v_{max},v_{max}]$.  We use the same velocity range for each
  velocity component and for all time.  (One could choose $v_{max} =
  c$ without doing a test simulation first, but for non-relativistic
  systems one would have many phase space cells with no particles,
  which leads to wasted memory and longer computational times for
  fixed velocity space bin size.)
\item Discretize velocity space by defining a velocity bin size
  $\Delta v$, which we choose to be the same in each direction in
  velocity space.  The velocity space bin size should be small enough
  to resolve typical velocity distribution functions, but large enough
  to preserve reasonably good statistics without many bins lacking
  particles, which leads to longer computational times.  If the
  velocity distributions in a system have known theoretical kinetic
  entropy values, a good way to determine $\Delta v$ is to compare the
  results using different $\Delta v$ with the predicted values, as we
  discuss further in Sec.~\ref{sec:deltav_dependence}.  We find that
  using a velocity space bin size comparable to the thermal speed is a
  good choice for the parameters of our simulation.

  Since $\Delta v$ determines the constant terms in
  Eqs.~(\ref{eq:stirling_f}), (\ref{eq:v_stirling}) and
  (\ref{eq:svelocity}), an absolute comparison of kinetic entropies of
  species with different $\Delta v$ would not be meaningful.  Instead,
  only relative changes to kinetic entropy should be used in such a
  case.  Therefore, for this initial study, we choose parameters so
  that the ion and electron thermal speeds are comparable, so we can
  use the same $\Delta v$ for both electrons and ions and be able to
  make direct comparisons.  For systems for which $v_{th,e}$ and
  $v_{th,i}$ are different, one should use different bin sizes for
  each species.  It is important to note that once the velocity space
  bin size for each species is set, it should be held fixed for the
  duration of the simulation and should be the same size for all grid
  cells.  These constraints are necessary to be able to compare
  kinetic entropies at different times and at different locations.
\item Choose a spatial bin size $\Delta x$.  In principle, this need
  not be the same as the grid scale $\Delta x$, but this is the most
  logical choice and what we employ here.  
 \item Cycle over every macro-particle and find the number density
   contribution to each spatial bin using the particle shape in the
   code, and increment its contribution to the number of
   macro-particles in the appropriate phase space bin based on the
   three components of the macro-particle's velocity.  The end result
   after counting all macro-particle contributions to every phase
   space bin is the total number of macro-particles in every bin
   $\mathcal{N}_{jk}$.  Recall, this typically is not an integer.
 \item If one wants to calculate the combinatorial Boltzmann entropy
   $\mathcal{S}$, then multiply $\mathcal{N}_{jk}$ in each bin by $a$
   to get $N_{jk}$.  As discussed in Appendix~\ref{appendix:macrovsreal},
   multiplying by $a$ at calculation time is not necessary for the
   continuous Boltzmann entropy $S$, but it would lead to an incorrect
   value of the combinatorial Boltzmann entropy $\mathcal{S}$.
\end{itemize}
Spatial cells at the boundary of a computational domain need to get
information from other processors for macro-particles in nearby cells
that contribute to $N_{jk}$.  This leads to an increase in run time;
for the present study, the ``base'' simulation takes $13\%$ more time
than the same simulation without calculating the kinetic entropy.  We
believe this performance could be improved, but leave that for future
work.

\subsection{Calculating Distribution Functions and Kinetic Entropies}
\label{appendix:calcdistfunc}

The distribution function $f(\vec{r}_j,\vec{v}_k)$ at bin $\vec{r}_j$
and $\vec{v}_k$ is immediately approximated from $N_{jk}$ using
Eq.~(\ref{eq:def_f}).  Once $f$ is obtained for all velocity space
bins in all spatial cells, the forms of continuous Boltzmann entropy
are readily calculated, such as Eq.~(\ref{eq:stirling_f}) for $S$, the
discretized version of Eq.~(\ref{eq:entropy_density}) for
$s(\vec{r}_j) = -k_B \sum_{k} (\Delta^3v) f(\vec{r}_j,\vec{v}_k) \ln
[f(\vec{r}_j,\vec{v}_k)]$, and Eq.~(\ref{eq:v_stirling}) for
$s_{\text{velocity}}$.  Note $S_{\text{position}}$ in
Eq.~(\ref{eq:space_stirling}) does not require the distribution
function.

To find the combinatorial Boltzmann entropy $\mathcal{S}$, use
Eq.~(\ref{eq:lngamma1}).  Since the $N_{jk}$ are not integers, the
factorial in Eq.~(\ref{eq:lngamma1}) needs to be reinterpreted using
the $\Gamma$ function for which $\Gamma(N+1) = N!$ for integer $N$
 \citep{Arfken95} as
%\begin{linenomath*}
\begin{equation}
  \mathcal{S}=k_B\left[\ln[\Gamma(N+1)]-\sum_{j,k}\ln[\Gamma(N_{jk}+1)]\right].
  \label{eq:lngamma}
\end{equation}
%\end{linenomath*}
Note that $N_{jk}$ need not be large in every cell, so the non-integer
part should not be ignored.  Fortuitously, many programming languages
contain an intrinsic function for $\ln[\Gamma(x)]$, so the calculation
is efficient and there are no issues with performing this calculation
for large argument [while calculating $\Gamma(N+1)$ separately would
  lead to numerical problems for large arguments].  A similar
calculation can be used to get the combinatorial Boltzmann entropy for
position and velocity space from Eqs.~(\ref{eq:ent_space}) and
(\ref{eq:ent_vel}), respectively.

\subsection{Merits of Combinatorial vs.~Continuous Boltzmann Entropy}
\label{appendix:calcrealunits}

We close this section with a discussion of the relative merits between
the combinatorial $\mathcal{S}$ and continuous $S$ Boltzmann kinetic
entropies.  Three advantages of the combinatorial Boltzmann entropy
are that it is the most accurate form of kinetic entropy (it does not
rely on assuming $N_{jk} \gg 1$), it is automatically in appropriate
units, and the intrinsic {\sc lngamma} function in many coding
languages makes the calculations efficient and more importantly can be
calculated for large argument, whereas a direct calculation taking the
factorial of a large number is not possible.  A drawback of the
combinatorial Boltzmann entropy is that the value of $a$, describing
the number of actual particles per macro-particle, must be included
from the beginning in the calculation.  Thus, if one wants to see how
the combinatorial Boltzmann entropy changes between two different
values of $a$, one must redo the calculation of kinetic entropy with a
different $a$ value.

The continuous Boltzmann entropy $S$ has the advantage that one does
not need to specify $a$ at run-time.  Therefore, finding the kinetic
entropy for the same simulation but with a different $a$ is trivial
and does not require redoing the calculation.  A disadvantage of the
continuous Boltzmann entropy is that one has to make sure that $a$ is
large enough that the errors in Stirling's approximation are small,
which is discussed further in Sec.~\ref{sec:a_dependence}. Another
disadvantage is that the results are not in appropriate units because
the argument of the natural logarithm in $S = -k_B \int d^3r d^3v f
\ln f$ from Eq.~(\ref{eq:stirling_f2}) is not dimensionless.  This
comes about because a term is dropped from Eq.~(\ref{eq:stirling_f}),
and the dropped term contains information about the units inside the
natural log. So, to convert the simulated continuous Boltzmann entropy
to real units for comparison to observations or experiments, one must
either (a) include the dropped term or (b) choose a reference value of
continuous Boltzmann entropy at a particular location and time and
present all values as a change in kinetic entropy relative to that
reference.  This enforces that the quantities have the appropriate
units.

%\bibliography{gcrbib-1}

%merlin.mbs aipnum4-1.bst 2010-07-25 4.21a (PWD, AO, DPC) hacked
%Control: key (0)
%Control: author (8) initials jnrlst
%Control: editor formatted (1) identically to author
%Control: production of article title (0) allowed
%Control: page (1) range
%Control: year (1) truncated
%Control: production of eprint (0) enabled
%

\end{document}